\documentclass[preprint]{aastex}

\usepackage{graphicx}
\usepackage{amssymb}
\usepackage{amsmath}
\usepackage{epstopdf}
\usepackage{natbib}
\usepackage{color}
\usepackage{hyperref}
\hypersetup{colorlinks=true, linkcolor=blue, citecolor=blue,urlcolor=blue}
\usepackage[all]{hypcap}
\usepackage[flushleft]{threeparttable}
\usepackage{lineno}

\def\eqlbl#1{\label{eq:#1}}
\def\eqref#1{(\ref{eq:#1})}

\shorttitle{Development of Habitable Climates}
\shortauthors{Brad Foley}

\begin{document}


\title{The Role of Plate Tectonic-Climate Coupling and Exposed Land Area in the Development of Habitable Climates on Rocky Planets}


\author{Bradford J. Foley \altaffilmark{1} }
\affil{Department of Terrestrial Magnetism, Carnegie Institution for Science, Washington, DC 20015}


\altaffiltext{1}{email: bradford.j.foley@gmail.com}



\begin{abstract} 

The long-term carbon cycle is vital for maintaining liquid water oceans on rocky planets due to the negative climate feedbacks involved in silicate weathering. Plate tectonics plays a crucial role in driving the long-term carbon cycle because it is responsible for CO$_2$ degassing at ridges and arcs, the return of CO$_2$ to the mantle through subduction, and supplying fresh, weatherable rock to the surface via uplift and orogeny. However, the presence of plate tectonics itself may depend on climate according to recent geodynamical studies showing that cool surface temperatures are important for maintaining vigorous plate tectonics. Using a simple carbon cycle model, I show that the negative climate feedbacks inherent in the long-term carbon cycle are uninhibited by climate's effect on plate tectonics. Furthermore, initial atmospheric CO$_2$ conditions do not impact the final climate state reached when the carbon cycle comes to equilibrium, as long as liquid water is present and silicate weathering can occur. Thus an initially hot, CO$_2$ rich atmosphere does not prevent the development of a temperate climate and plate tectonics on a planet. However, globally supply-limited weathering does prevent the development of temperate climates on planets with small subaerial land areas and large total CO$_2$ budgets because supply-limited weathering lacks stabilizing climate feedbacks. Planets in the supply-limited regime may become inhospitable for life and could experience significant water loss. Supply-limited weathering is less likely on plate tectonic planets, because plate tectonics promotes high erosion rates and thus a greater supply of bedrock to the surface.



\end{abstract}


\keywords{astrobiology --- planets and satellites: physical evolution --- planets and satellites: terrestrial planets}

\section{Introduction}
\label{sec:intro}

One of the biggest questions surrounding rocky extra-solar planets is whether they are potentially habitable for life. Water is thought to be necessary for the development of life, so a habitable planet is typically defined as one that is able to sustain liquid water oceans \citep[e.g.][]{Kasting2003}. In order to be habitable a planet must lie within the habitable zone, the range of orbital distances where water can exist as a stable phase on a rocky planet's surface \citep[e.g.][]{Hart1978,Hart1979,Kasting1993,Franck2000}, and it must have accreted enough water to produce oceans. However, lying within the habitable zone does not guarantee that a planet, even one that accreted a sufficient supply of H$_2$O, will have liquid water oceans, or be able to maintain oceans for a significant portion of its history. The abundance of greenhouse gases in a planet's atmosphere is also critical. A planet with a strong greenhouse effect could be hot enough to enter a moist greenhouse state, where dissociation of water in the upper atmosphere and hydrogen escape to space leads to rapid water loss \citep[e.g.][]{Kasting1988,Abbot2012}. Furthermore, stellar evolution exerts a major influence on planetary climate that must be balanced by atmospheric greenhouse gas abundances \citep{Kasting1989}. Stars increase in luminosity as they age, so early in a planet's history high greenhouse gas concentrations are needed to avoid a snowball climate \citep{Sagan1972,Newman1977,Gough1981}, and late in a planet's history low greenhouse gas concentrations are needed to keep surface temperatures below the moist greenhouse limit, where rapid water loss occurs \citep{Kasting1988}.  The same effects apply to planets at different orbital distances. Planets closer to the inner edge of the habitable zone need a lower greenhouse effect to maintain oceans, while those closer to the outer edge need a strong greenhouse effect to prevent global glaciation. 

A mechanism capable of regulating the strength of the atmospheric greenhouse effect, and thus counteracting the spatial and temporal variations in solar luminosity described above, exists via the long-term carbon cycle, which controls atmospheric CO$_2$ concentrations \citep[e.g.][]{Walker1981,Kasting1993,Berner2004}. The long-term carbon cycle refers to the cycling of CO$_2$ between the atmosphere and ocean, carbonate rocks on the seafloor, and the mantle. Weathering of silicate minerals on continents and in the oceanic crust draws CO$_2$ out of the atmosphere and ocean, depositing it on ocean plates in the form of carbonate rocks. CO$_2$ resides on the seafloor until it is subducted, where some portion of the carbon reaches the deep mantle, and the rest is returned to the atmosphere through metamorphic degassing and arc volcanism. The carbon in the mantle eventually degasses back to the atmosphere and ocean at mid-ocean ridges, completing the cycle. The long-term carbon cycle stabilizes planetary climate due to the temperature sensitivity of silicate weathering. At high temperatures weathering rates increase, and more rapid weathering draws CO$_2$ out of the atmosphere, cooling the climate. Meanwhile at low temperatures weathering rates decrease (or cease entirely in the case of an ice covered planet), and sluggish weathering allows volcanic degassing to build CO$_2$ back up in the atmosphere, warming the climate \citep[e.g.][]{Berner2004}. The same feedback buffers climate against changes in solar luminosity. When luminosity is low, slower weathering rates allow large quantities of CO$_2$ to buildup in the atmosphere, warming the climate, and when luminosity is high rapid weathering draws down atmospheric CO$_2$, cooling the climate \citep[e.g.][]{Walker1981}. As a result, an active global carbon cycle is thought to be essential for the  long-term maintenance of liquid water oceans on rocky planets within the habitable zone \citep{Kasting2003}. 
 
Plate tectonics plays a vital role in the operation of the long-term carbon cycle. Plate tectonics drives volcanism at ridges and arcs, the major sources of atmospheric CO$_2$, and facilitates silicate weathering, the primary sink of atmospheric CO$_2$, by providing a continuous supply of fresh, weatherable rock (both on continents and on the seafloor) through orogeny and volcanic resurfacing. Therefore plate tectonics is often thought to be necessary for the operation of the long-term carbon cycle, and thus crucial for allowing planets throughout the habitable zone to sustain liquid water oceans over geologic timescales \citep{Gonzalez2001,Kasting2003} (though it is unknown whether some form of carbon cycling and climate stabilization can occur on non-plate-tectonic planets). Studies of carbon cycling and planetary habitability typically assume that plate tectonics and climate are independent; the presence of plate tectonics is imposed by the modeler, and the plate speed is often taken as a free parameter \citep{Driscoll2013}, or as solely a function of mantle temperature \citep[e.g.][]{TajikaMatsui1992,Franck1999,Sleep2001b}. However, recent geodynamical studies have found that climate and plate tectonics might be linked. Cool surface temperatures promote plate tectonics by suppressing grain-growth, allowing weak plate boundaries to form through grainsize reduction \citep{Landuyt2009a,Foley2012,br2014}, and by increasing mantle convective stresses, such that convective forces can more easily exceed the lithosphere's intrinsic strength \citep{Lenardic2008}. Plate speed may even depend on surface temperature, as higher surface temperatures lead to stronger lithospheric shear zones via rapid grain-growth, and these stronger shear zones provide a larger resistance to plate motions \citep{Foley2013_scaling,Foley2014_initiation}. Although there is uncertainty over the applicability of these geodynamical studies to the tectonics of terrestrial planets, as the mechanism responsible for plate tectonics on Earth is still not fully known \citep{Tackley2000,berco2003}, the implications of a fully coupled climate-mantle system (where the presence and vigor of plate tectonics depends on surface temperature) for planetary habitability are worth exploring.  

As stated in the previous paragraph, one of the key roles plate tectonics plays in the global carbon cycle is providing a supply of fresh rock at the surface. Without a sufficient supply of fresh rock continental weathering can enter a supply-limited regime. In the supply-limited regime the weathering reaction runs to completion in the regolith, and weathering can only continue by bringing unaltered bedrock into the weathering zone through physical erosion. As a result, supply-limited weathering depends on the physical erosion rate, rather than the kinetics of the weathering reaction \citep[e.g.][]{Stallard1983,Edmond1995,Kump2000,Riebe2004,West2005,Hilley2010}. If continental weathering on a planet becomes globally supply-limited, it would no longer depend on atmospheric CO$_2$ and surface temperature, and the climate stabilization provided by the long-term carbon cycle could be lost \citep{West2012}. Given the important link between plate tectonics and the supply of weatherable rock, including a treatment for supply-limited weathering is crucial for studying the coupling between plate tectonics, climate, and planetary habitability. Furthermore, supply-limited weathering could be especially important on planets with small amounts of subaerial land (i.e. waterworlds), where the supply of fresh rock at the surface would naturally be low. Yet supply-limited weathering is often ignored in studies of exo-planet habitability, which, so far, have optimistically concluded that climate stabilization via the long-term carbon cycle can still be maintained on planets where up to 99 \% of the surface is covered by oceans \citep[e.g.][]{Abbot2012}.

Therefore in this study I test the ability of the long-term carbon cycle to regulate planetary climate, and thus allow hospitable surface conditions to persist over geologic time, for a fully coupled climate-mantle system that includes the supply limit to weathering. In particular, I assess whether the negative climate feedbacks inherent in the long-term carbon cycle are preserved when the plate speed depends on surface temperature as suggested by \cite{Foley2013_scaling} and \cite{Foley2014_initiation}, and the influence initial atmospheric CO$_2$ concentrations have on the final steady-state climate reached on a planet. Finally, I determine whether planets with small land areas can maintain a carbon cycle with negative climate feedbacks when supply-limited weathering is considered. This study only considers Earth-like planets, that is planets whose size and bulk composition are similar to the Earth's, as these type of planets are most easily modeled based on our current understanding of climate, tectonics, and carbon cycling on Earth (the possible influences of planet size are discussed in \S \ref{sec:size}).  

This study also focus solely on CO$_2$ cycling, which is thought to exert the leading order control on the climate evolution of rocky planets \citep[e.g.][]{Kasting2003,Berner2004}. One reason for the prominence of CO$_2$ is that it is a major component of the gases released by terrestrial magmatism on Earth today, likely because the upper mantle became oxidized during, or soon after, core formation \citep[e.g.][]{Catling2005,Wade2005,Frost2008}. Thus mantle degassing is a major source of atmospheric CO$_2$, and without weathering to act as a sink for this CO$_2$, Earth's climate would be inhospitable to life. Carbon dioxide degassing has likely been active for the majority of Earths' history, because core formation on the Earth was rapid, due to the high interior temperatures reached during accretion \citep{Kleine2002}. Geochemical data indicate that the upper mantle was oxidized by 3.9-3.5 Ga \citep{Delano2001,Frost2008_rev}, and possibly as early as $\approx 4.4$ Ga \citep{Trail2011}. Early oxidation of the mantle during core formation is thought to be a general feature of terrestrial planet formation, at least for planets as large or larger than the Earth \citep{Halliday2007}, so CO$_2$ is likely to be a significant component of the atmospheres of rocky exo-planets as well. Although carbon dioxide is not the only greenhouse gas that can play a key role in planetary habitability, methane in particular may be important for solving the faint young sun problem \citep{Pavlov2000,Haqq2008,Feulner2010,Charnay2013,Wolf2013}, the ability of silicate weathering to regulate atmospheric CO$_2$ is likely to be crucial for the habitability of rocky planets.   

The paper is organized in the following manner: \S \ref{sec:model} describes the global carbon cycle model used in this study, with the weathering formulation that includes supply-limited weathering laid out in \S \ref{sec:weathering_flux}, and the dependence of plate speed on surface temperature described in \S \ref{sec:climate_model}; results are shown in \S \ref{sec:results}, with the results of models testing the stability of the fully coupled climate-mantle system given in \S \ref{sec:stability}, and the influence of supply-limited weathering on climate stabilization for planets with small land areas shown in \S \ref{sec:fland}; finally, results are discussed in \S \ref{sec:discussion} and the main conclusions are summarized in \S \ref{sec:conclusions}.

\section{Model Setup} 
\label{sec:model}
\begin{figure}
\includegraphics[scale = 0.75]{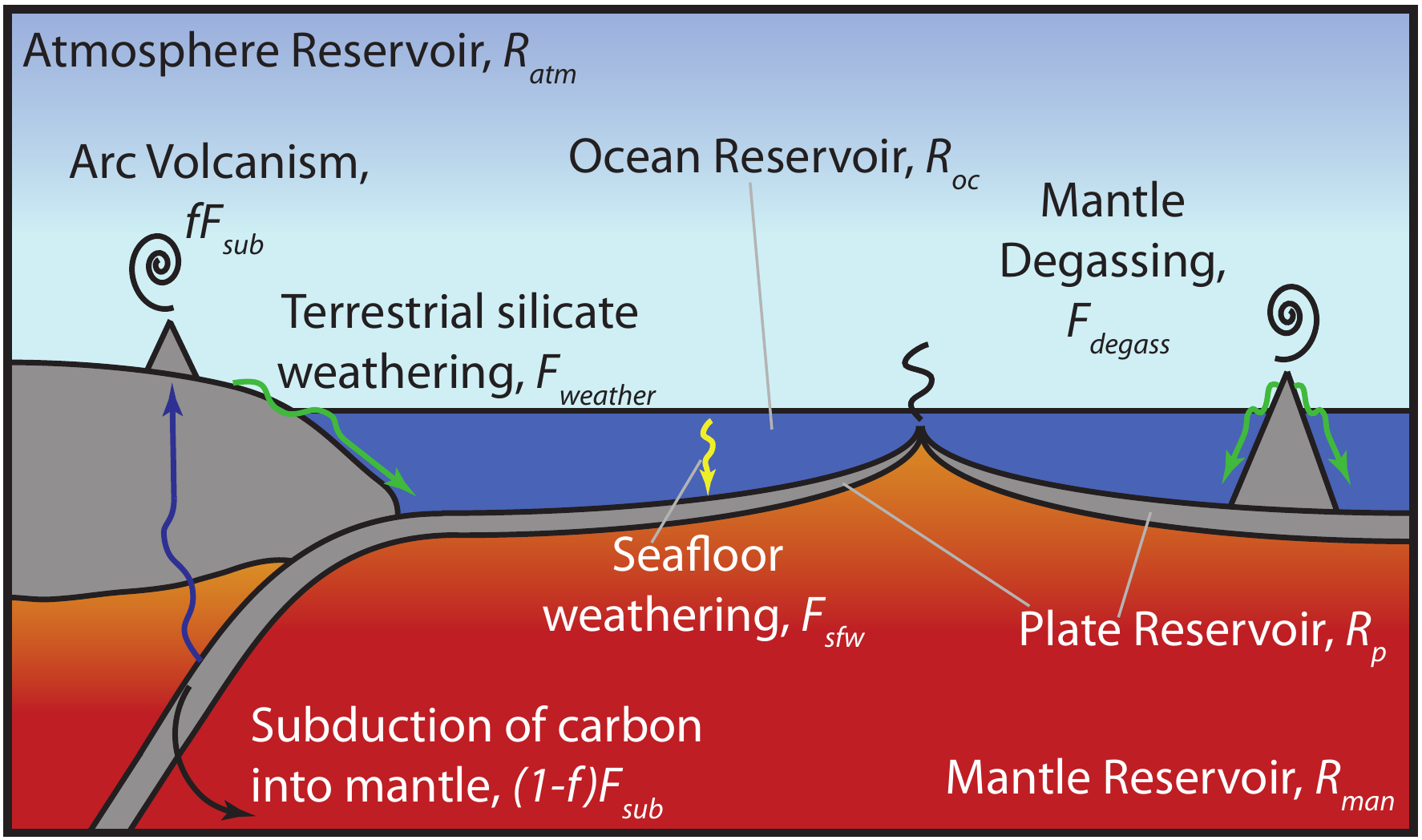}
\caption{\label {fig:c_cycle_model} Schematic diagram of the global carbon cycle model used in this study.}  
\end{figure}

Global carbon cycle models of varying complexity have been used to study the long-term evolution of the Earth and other planets \citep[e.g.][]{Walker1981,TajikaMatsui1990,TajikaMatsui1992,Franck1999,Sleep2001b,Sleep2001a,Abbot2012,Driscoll2013}, as well as for more detailed studies of Neoproterozoic and Phanerozoic climate evolution \citep[e.g.][]{Berner1983,Volk1987,Berner1994,Berner2004,Tajika1998,Mills2011}. In this study a simple model is used to focus on the first order effects of both climate-tectonic coupling and exposed land area on climate stabilization via the global carbon cycle.  Carbon is assumed to cycle between four reservoirs: the ocean crust (or plate) ($R_p$), the mantle ($R_{man}$), the atmosphere ($R_{atm}$), and the ocean ($R_{oc}$) (Figure \ref{fig:c_cycle_model}). A significant fraction of oceanic carbonates reside on continental shelves at the present day \citep[e.g.][]{Ronov1969}, and thus a separate continental reservoir is often included in global carbon cycle models \citep[e.g.][]{TajikaMatsui1990,Sleep2001b}. Here all carbonates are assumed to form on the seafloor, i.e. they are part of the ocean crust reservoir, as in \cite{Driscoll2013}. The separate continental reservoir is neglected because much of this study focuses on planets with small land fractions, where the large majority of carbonate formation would occur on the seafloor. Furthermore, neglecting the continental reservoir does not significantly alter the first order climate feedbacks or climate-tectonic coupling, because the functional form of the degassing function for continental carbon is similar to the mantle degassing function, in that both depend linearly on plate speed \citep{Sleep2001b}. 

Carbon cycles between the four reservoirs in this model as follows (Figure \ref{fig:c_cycle_model}): Exposed rock at the surface reacts with atmospheric CO$_2$ to form bicarbonate, magnesium, and calcium ions that travel to the oceans via rivers and groundwater. Once in the ocean, these ions form carbonate minerals on the seafloor, both in the form of sediments and via hydrothermal alteration of basalt (CO$_2$ dissolved in the oceans can also react directly with basalt and form carbonate minerals, regardless of whether silicate weathering is active) \citep[e.g.][]{Berner2004}. Thus weathering acts as a net flux of carbon from the atmosphere and ocean to the plate reservoir. Carbon leaves the plate reservoir when it is subducted at trenches. Here, a fraction of the carbonate minerals will devolatilize, returning carbon to the atmosphere via arc volcanism, and the remainder will be subducted into the deep mantle. Once in the mantle, carbon is mixed throughout by convection, and degasses to the atmosphere and ocean at mid-ocean ridges. Return of carbon to the atmosphere and ocean via mantle degassing closes the long-term carbon cycle. 

The carbon cycle is modeled using the following mass balance equations \citep[e.g.][]{TajikaMatsui1992,Sleep2001b,Driscoll2013}: 
\begin{equation}
\eqlbl{rp_eq}
\frac{\mathrm{d} R_p}{\mathrm{d}t} = \frac{F_{weather}}{2} + F_{sfw} - F_{sub}
\end{equation}
\begin{equation}
\eqlbl{rman_eq}
\frac{\mathrm{d} R_{man}}{\mathrm{d}t} = (1-f)F_{sub} - F_{degas}
\end{equation}
\begin{equation}
\eqlbl{raoc_eq}
\frac{\mathrm{d} (R_{atm} + R_{oc}) }{\mathrm{d}t} = F_{arc} + F_{degas} - \frac{F_{weather}}{2} - F_{sfw} 
\end{equation}
where $F_{weather}$ (the continental or terrestrial weathering flux) is the flux of carbon from the atmosphere to the plate resulting from silicate weathering on land, $F_{sfw}$ (the seafloor weathering flux) is the flux of carbon from the ocean to the plate resulting from hydrothermal alteration of basalt by CO$_2$ dissolved in seawater, $F_{sub}$ (the subduction flux) is the flux of carbon from the plate into the mantle, $F_{degas}$ (the degassing flux) is the flux of carbon from the mantle to the atmosphere and oceans via mid-ocean ridge volcanism, and $F_{arc}$ (the arc volcanic flux; $F_{arc} = f F_{sub}$, where $f$ is the fraction of subducted carbon that is degassed at arcs) is the flux of carbon to the atmosphere from devolatilization of the subducting slab (see Table \ref{tab_var} for a list of all model variables). The continental weathering flux is divided by two because half of the carbon initially drawdown by weathering on land is rereleased to the atmosphere when carbonates form on the seafloor \citep[e.g.][]{Berner1983}. In other words, $F_{weather}/2$ represents the net flux of CO$_2$ from the atmosphere and ocean reservoirs to the plate reservoir via continental weathering. The seafloor weathering flux only represents the alteration of basalt by CO$_2$ dissolved in the oceans; formation of carbonates in basalt by calcium and bicarbonate ions derived from silicate weathering on land are included in $F_{weather}$. Thus, equation \eqref{rp_eq} describes how carbon is added to the plate reservoir through continental and seafloor weathering, and leaves via subduction, equation \eqref{rman_eq} tracks the addition of carbon to the mantle reservoir via deep subduction of carbonates (i.e. $(1-f)F_{sub}$), and its removal by ridge degassing, and equation \eqref{raoc_eq} describes how arc volcanism and mantle degassing add carbon to the atmosphere and ocean, while silicate weathering removes it. This model also assumes that carbon is instantaneously mixed throughout the mantle after being subducted. This is a simplification because mantle mixing is not instantaneous, and can take up to 1 Gyr for whole mantle mixing in the present day Earth \citep[e.g.][]{Olson1984,Hoffman1985,Christensen1989,Kellogg1990}. However, as mantle mixing with realistic geometries and rheological effects is still a major topic of research \citep{Tackley_treatise}, incorporating the influence of a finite mixing time for mantle carbon is left for future studies. 

As in \cite{Sleep2001b}, the atmosphere and ocean reservoirs are grouped together. The partitioning of carbon between the atmosphere and ocean is controlled by the solubility of CO$_2$ in the oceans. These two reservoirs equilibrate on a very short timescale compared to the timescale over which the other reservoirs evolve. Thus, equilibration between the atmosphere and ocean is assumed to take place instantaneously, and the combined atmosphere plus ocean carbon reservoir is partitioned into $R_{atm}$ and $R_{oc}$ using Henry's Law at each timestep. Henry's Law is given by $P_{CO_2} = k_c x_c$, where $P_{CO_2}$ is the partial pressure of atmospheric CO$_2$, $k_c$ is the solubility and $x_c$ is the mole fraction of CO$_2$ in the oceans; $x_c = R_{oc}/(M_{H_2O}  + R_{oc})$ where $M_{H_2O}$ is the number of moles of H$_2$O in an ocean mass of water. The temperature dependence of the solubility, $k_c$, is ignored in this study for simplicity; a $k_c$ representative of warm greenhouse conditions, where less CO$_2$ can be dissolved in the oceans, is chosen and used in the majority of the models (see Table \ref{tab_param} for a list of all model parameters and their assumed baseline value, and Table \ref{tab_cons} for a list of all physical constants used in this study). Variations in $k_c$ are considered in \S \ref{sec:param_test} and found to have no significant influence on the results. The atmospheric CO$_2$ reservoir can be related to the partial pressure of atmospheric CO$_2$ ($P_{CO_2}$) through 
\begin{equation}
\eqlbl{pco2}
P_{CO_2} \approx \frac{R_{atm} \bar{m}_{CO_2} g}{A_{earth}} , 
\end{equation}
where $\bar{m}_{CO_2}$ is the molar mass of CO$_2$, $g$ is acceleration due to gravity, and $A_{earth}$ is the surface area of the Earth. This relation is only an approximation, because relating the moles of atmospheric CO$_2$ to $P_{CO_2}$ depends on the composition of the rest of the atmosphere. If CO$_2$ is the dominant component, then equation \eqref{pco2} is exact; if N$_2$ or H$_2$O are the dominant components in the atmosphere, then equation \eqref{pco2} will overestimate $P_{CO_2}$ by approximately a factor of 2. However, the impact of this error on the overall results of the study is negligible; test cases assuming that CO$_2$ is a minor atmospheric component were found to be nearly identical to the results presented in the remainder of this paper, all of which use equation \eqref{pco2}.

\begin{table}
\caption{Table of variables. }
\label{tab_var}
\begin{tabular}{l l c}
\hline
Symbol & Definition and units & Equation \\ \hline
$R_p$ & Ocean plate carbon reservoir (mol) & \eqref{rp_eq} \\
$R_{man}$ & Mantle carbon reservoir (mol) & \eqref{rman_eq} \\
$R_{atm}$ & Atmosphere carbon reservoir (mol) & \eqref{raoc_eq} \\
$R_{oc}$ & Ocean carbon reservoir (mol) & \eqref{raoc_eq} \\ 
$F_{sub}$ & Subduction flux (mol Ma$^{-1}$) & \eqref{fsub} \\ 
$F_{degas}$ & Degassing flux (mol Ma$^{-1}$) & \eqref{fd} \\  
$F_{weather}$ & Continental weathering flux (mol Ma$^{-1}$) & \eqref{weather} \\  
$F_{w_s}$ & Supply-limited weathering flux (mol Ma$^{-1}$) & \eqref{total_weathering_flux} \\  
$F_{w_k}$ & Kinetically-limited weathering flux (mol Ma$^{-1}$) & \eqref{my_fw} \\ 
$F_{sfw}$ & Seafloor weathering flux (mol Ma$^{-1}$) & \eqref{fsfw} \\  
$P_{CO_2}$ & Partial pressure of atmospheric CO$_2$ (Pa) & \eqref{pco2} \\  
$A_p$ & Area of ocean plates ($A_{Earth}f_{land}$, km$^2$) & \eqref{fsub} \\
$f_{land}$ & Land fraction & \eqref{convert_west_2012} \\
$P_{sat}$ & Saturation vapor pressure (Pa) & \eqref{psat} \\  
$T$ & Surface temperature (K) & \eqref{temp} \\  
$T_e$ & Effective temperature (K) & \eqref{te} \\  
$S$ & Solar irradiance (W m$^{-2}$) & \eqref{te} \\  
$v$ & Plate speed (cm yr$^{-1}$ or m Ma$^{-1}$) & \eqref{plate_speed} \\  
$C_{tot}$ & Total planetary CO$_2$ budget (mol) & \eqref{C_tot} \\
\end{tabular}      
\end{table}

\subsection{Subduction, Arc, Degassing, and Seafloor Weathering Fluxes}

The subduction flux, $F_{sub}$, is given by the product of the area density of carbon on the plate, $R_p/A_{p}$ (where $A_p$ is the area of oceanic plates), the plate speed, $v$, and the length of trenches, $L$, 
\begin{equation}
\eqlbl{fsub}
F_{sub} = \frac{R_p v L}{A_p} .
\end{equation} 
The arc flux is then $F_{arc} = f F_{sub}$, and the flux of carbon into the deep mantle is $(1-f) F_{sub}$. The fraction of subducting carbon that degasses through arc volcanism on the present day Earth is not well constrained, with estimates typically ranging from 25 \% to 70 \% \citep[e.g.][]{Sleep2001b,Dasgupta2010,Ague2014}, though a vary small $f<0.1$ can not be ruled out \citep{Kelemen2015}. Thus $f = 0.5$ is chosen as a baseline; variations in $f$ are considered in \S \ref{sec:param_test}, and do not significantly influence the main results of this study (see also \S \ref{sec:uncertainties}). 

The degassing flux is given by the flux of upwelling mantle into the melting region beneath the ridge, multiplied by the concentration of carbon in the mantle and the fraction of this carbon that degasses \citep{TajikaMatsui1992}. The flux of mantle into the melting zone must balance the flux of mantle leaving this region, and is thus given by $2vLd_{melt}$, where $d_{melt}$ is the depth where mid-ocean ridge melting begins and the length of ridges is assumed to be equal to the length of trenches. The degassing flux is then 
\begin{equation} 
\eqlbl{fd}
F_{degas} = f_d \frac{R_{man}}{V_{man}} 2 v L d_{melt} 
\end{equation} 
where $f_d$ is the fraction of upwelling mantle that degasses and $V_{mantle}$ is the volume of the mantle. The current day length of mid-ocean ridges is $\approx 6 \times 10^4$ km \citep[e.g.][]{Fowler_book}; this value is assumed to be constant and used for all models. The degassing fraction, $f_d$, is fixed to reproduce the present day Earth's degassing flux and atmospheric CO$_2$ content, which gives $f_d \approx 0.32$, similar to the estimate of \cite{TajikaMatsui1992}.  

The seafloor weathering flux, in particular whether it has a strong climate feedback or not, is poorly understood. Drill cores of oceanic crust show that low temperature (0-60$^{\circ}$ C) off-axis hydrothermal alteration of basalt is a significant CO$_2$ sink \citep{Staudigel1989,Alt1999,Gillis2011} and laboratory measurements indicate that the rate of CO$_2$ uptake via basalt alteration increases with both increasing temperature and $P_{CO_2}$ \citep{Brady1997}. However, it is not clear if the temperature of the seawater circulating through crustal basalt during hydrothermal alteration is dictated by the geothermal heat flow \citep{Alt1999} or by the ocean bottom water temperature \citep{Gillis2011,Coogan2013,Coogan2015}, which is related to climate. Furthermore, seafloor weathering can also become ``supply-limited," if all of the basalt accessible to hydrothermal fluids is completely carbonated \citep{Sleep2001a}; in this case only the creation of new seafloor at mid-ocean ridges allows further seafloor weathering. Thus even with a direct temperature feedback there are limits to the amount of CO$_2$ that can be sequestered in the ocean crust. In this study both the direct temperature dependence of seafloor weathering and its supply-limit are neglected due to the large uncertainties associated with these effects; this also results in a simpler model that can be more completely understood and analyzed. In \S \ref{sec:sfwt} I show that including a direct temperature feedback and a supply-limit to seafloor weathering does not significantly change the results of this study, therefore justifying the exclusion of these two effects, as long as complete basalt carbonation cannot penetrate to great depths within the crust. Physically, the simplified seafloor weathering flux used in this study can be interpreted as assuming that crustal heat flow sets the temperature of the water reacting with basaltic crust \citep{Abbot2012}. The seafloor weathering flux therefore follows after \cite{Sleep2001b} and \cite{Mills2014} as,
\begin{equation}
\eqlbl{fsfw}
F_{sfw} = F_{sfw}^* \left(\frac{v}{v^*}\right) \left(\frac{P_{CO_2}}{P_{CO_2}^*}\right)^{\alpha}, 
\end{equation}
where stars denote present day quantities, $\alpha$ describes the dependence of basalt carbonation on atmospheric CO$_2$, and $(v/v^*)$ represents the effect of spreading rate, with a constant ridge length, $L$. Spreading rate sets the rate at which fresh, weatherable basalt is created, and the $P_{CO_2}$ dependence is based on \cite{Brady1997}, who find $\alpha \approx 0.23$; this is rounded up to 0.25 for the baseline value of $\alpha$ used in this study. The current day seafloor weathering flux, $F_{sfw}^*$, is set to $1.75 \times 10^{12}$ mol/yr \citep{Mills2014}.   

\subsection{Continental Weathering Flux} 
\label{sec:weathering_flux}

The continental weathering flux typically used in global carbon cycle models is based on the kinetics of the weathering reaction between CO$_2$ and silicate rocks, with parameterizations for additional effects, such as increases in weathering rates brought about by increases in runoff or tectonic activity \citep[e.g.][]{Walker1981,TajikaMatsui1992,Sleep2001b,Berner2004,Driscoll2013}. However, the weathering flux will only be determined by the reaction kinetics when the weathering reaction is the rate limiting step (i.e. when weathering is kinetically-limited). When weathering rates are high, or physical erosion rates are low, weathering can be limited by the supply of fresh rock to the surface \citep[e.g.][]{Stallard1983,Edmond1995,Kump2000,Riebe2004,West2005}. A more general weathering function, that incorporates supply-limited weathering, is given by \citep{Gabet2009,Hilley2010,West2012},
\begin{equation}
\eqlbl{west_2012}
W = \varepsilon \chi_{cc} \left[1 - \exp{\left(-R \frac{z}{\varepsilon} \right)} \right] 
\end{equation}   
where $W$ is weathering rate (with dimensions of mass area$^{-1}$ time$^{-1}$), $\varepsilon$ is the total denudation rate (mass area$^{-1}$ time$^{-1}$), $\chi_{cc}$ is the fraction of reactable cations in the continental crust and is unitless, $R$ is the rate of the silicate weathering reaction (time$^{-1}$), and $z$ is the effective depth of the weathering zone (mass area$^{-1}$). The weathering rate, $W$, can be converted into the global weathering flux, $F_{weather}$, by integrating over the area of exposed land, and dividing by the molar mass of reactable cations in the bedrock. Thus, 
\begin{equation}
\eqlbl{convert_west_2012}
F_{weather} = \frac{WA_{earth}f_{land}}{\bar{m}_{cc}}
\end{equation}   
where $f_{land}$ is the fraction of exposed land (i.e. the area of exposed land divided by the surface area of the Earth), and $\bar{m}_{cc}$ is the average molar mass of reactable elements in the continental crust. Combining equations \eqref{west_2012} and \eqref{convert_west_2012}, 
\begin{equation}
\eqlbl{total_weathering_flux}
F_{weather} =  \frac{A_{earth}f_{land}E \rho_r \chi_{cc}}{\bar{m}_{cc}} \left(1 - \exp{\left(-R \frac{z}{\varepsilon}\right)} \right) = F_{w_s} \left(1 - \exp{\left(-R \frac{z}{\varepsilon}\right)} \right)
\end{equation}  
where $E$ is the physical erosion rate (length time$^{-1}$), $\rho_r$ is the density of the regolith, and $F_{w_s} = (A_{earth}f_{land} E \chi_{cc} \rho_r)/\bar{m}_{cc} $ is the supply limit to weathering. Physical erosion rates vary by a wide range on Earth, depending on relief, climate, lithology, and other factors. In this study, a full model of how erosion varies in response to these factors is not attempted. Instead, a maximum erosion rate, $E_{max}$, representing the upper bound on the globally averaged erosion rate on a planet, is chosen. This maximum erosion rate, along with the area of exposed land, then sets the global supply limit to weathering, $F_{w_s}$. $E_{max}$ is poorly constrained, so a wide range of maximum erosion rates are tested in \S \ref{sec:param_test}, and the likely dependence of $E_{max}$ on tectonic activity, and the implications this has for weathering and climate regulation on stagnant lid planets, is discussed in \S \ref{sec:pt_weathering}.    

The reaction rate, $R$, is determined from the typical weathering flux equations used in previous global carbon cycle models \citep[e.g.][]{Walker1981,TajikaMatsui1992,Sleep2001b,Berner2004,Driscoll2013}; in this way the total weathering flux will mimic previous models when weathering is kinetically-limited, and then plateau when weathering rates are high enough, or erosion rates low enough, to reach the supply limit.  Following \cite{Driscoll2013}, the weathering flux for kinetically-limited weathering is 
\begin{equation} 
\eqlbl{my_fw}
F_{w_k} = F_w^* \left(\frac{P_{CO_2}}{P_{CO_2}^*} \right)^{\beta} \left(\frac{P_{sat}}{P_{sat}^*} \right)^{a} \exp{\left(\frac{E_a}{R_g} \left(\frac{1}{T^*} - \frac{1}{T}  \right)\right)} \left(\frac{f_{land}}{f_{land}^*} \right)
\end{equation}
where $F_w^*$ is the present day weathering flux ($F_w^* \approx 12 \times 10^{12}$ mol/yr \citep{Gaillardet1999}), $P_{sat}$ is the saturation vapor pressure, $\beta$ and $a$ are constants, $E_a$ is the activation energy of the weathering reaction, $R_g$ is the universal gas constant, $T$ is the surface temperature, and starred quantities represent present day values.  The weathering flux increases with increasing $P_{CO_2}$ or $T$ because both factors increase the rate of the weathering reaction \citep{Berner2004}, and the saturation vapor pressure term models the variation in runoff as surface temperature changes \citep{Driscoll2013}. The saturation vapor pressure is \citep[e.g.][]{Kasting1984,Nakajima1992}
\begin{equation}
\eqlbl{psat}
P_{sat} = {P_{sat}}_0 \exp{\left[-\frac{\bar{m}_w L_w}{R_g} \left(\frac{1}{T} - \frac{1}{{T_{sat}}_0} \right) \right]}
\end{equation}
where ${P_{sat}}_0$ is the reference saturation vapor pressure, at reference temperature ${T_{sat}}_0$, $\bar{m}_w$ is the molar mass of water, and $L_w$ is the latent heat of water (see Tables \ref{tab_param} \& \ref{tab_cons}). 

\begin{figure}
\includegraphics[scale = 0.75]{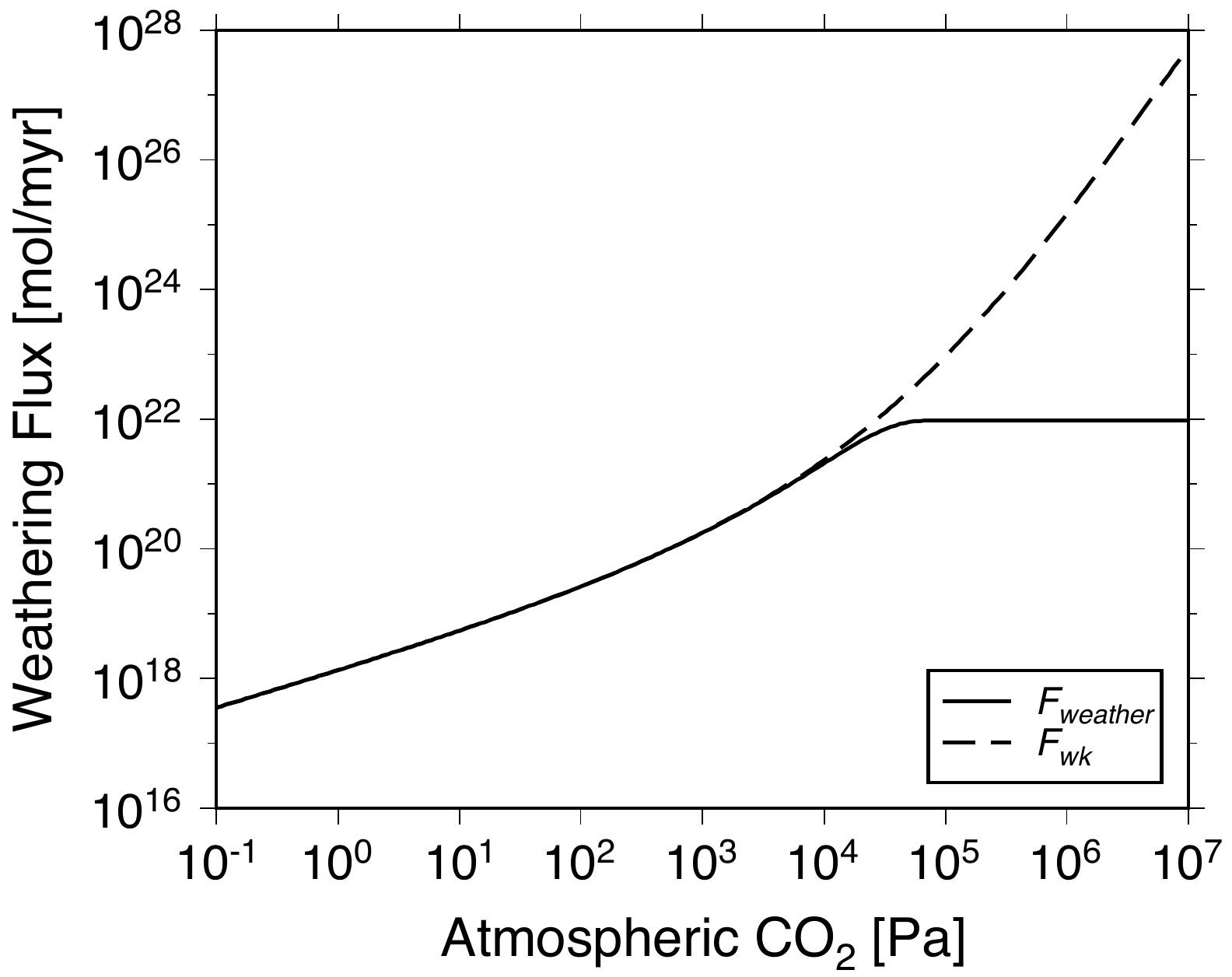}
\caption{\label {fig:f_weather_comp} Comparison between the total weathering flux, including supply-limited weathering, $F_{weather}$ (solid line), and the kinetically-limited weathering flux, $F_{w_k}$ (dashed line). Surface temperature depends on $P_{CO_2}$ based on equation \eqref{temp} and saturation vapor pressure depends on surface temperature as given by equation \eqref{psat}. A maximum erosion rate of 10 mm/yr and Earth-like land fraction of 0.3 were assumed, and the other parameters are listed in Tables \ref{tab_param} \& \ref{tab_cons}.}  
\end{figure}

Assuming that the kinetically-limited global weathering flux is equal to the reaction rate, $R$, multiplied by the total amount of reactable cations within the weatherable bedrock,  
\begin{equation} 
\eqlbl{convert_kinetic}
F_{w_k} = R \left(\frac{h_w A_{earth} f_{land} \rho_r \chi_{cc}}{\bar{m}_{cc}}\right) = R \left(\frac{z A_{earth} f_{land} \chi_{cc}}{\bar{m}_{cc}}\right) 
\end{equation}
where $h_w$ is the thickness of the weathering zone, which can be replaced by $z/\rho_r$ \citep{West2012}.  Combining equations \eqref{total_weathering_flux}, \eqref{my_fw}, and \eqref{convert_kinetic}, the definition of $F_{w_s}$, and using $\varepsilon = E_{max}\rho_r$,  
\begin{equation}
\eqlbl{weather}
 F_{weather} =  F_{w_s} \left \{ 1 - \exp{\left[-\frac{F_w^* f_{land}}{F_{w_s} f_{land}^*} \left(\frac{P_{CO_2}}{P_{CO_2}^*} \right)^{\beta} \left(\frac{P_{sat}}{P_{sat}^*} \right)^{a} \exp{\left(\frac{E_a}{R_g} \left(\frac{1}{T^*} - \frac{1}{T}  \right)\right)} \right] }\right \} .
\end{equation}
Inverting for the reaction rate using a weathering function normalized to the present day Earth (equation \eqref{convert_kinetic}) implicitly assumes that weathering on the present day Earth is entirely kinetically controlled. While the breakdown of the present day global weathering flux between supply-limited and kinetically-limited weathering is not well constrained, many estimates find that kinetically-limited weathering is the dominant contributor \citep{Kump2000,West2005}. Thus assuming Earth's present day weathering flux is kinetically controlled is a reasonable first order approximation. 

The weathering flux, $F_{weather}$, is shown in Figure \ref{fig:f_weather_comp}, plotted against $F_{w_k}$, i.e. the typical weathering flux used in global carbon cycle models that do not treat supply-limited weathering. $F_{weather}$ is identical to $F_{w_k}$ at low $P_{CO_2}$ (corresponding to low surface temperatures and low weathering rates), and diverges from $F_{w_k}$ at high $P_{CO_2}$ when weathering becomes supply-limited; here $F_{weather}$ plateaus at $F_{w_s}$, the global supply limit to weathering, determined by the maximum erosion rate and area of exposed land at the surface.        

\subsection{Climate and Plate Tectonic Models} 
\label{sec:climate_model}

\begin{figure}
\includegraphics[scale = 0.75]{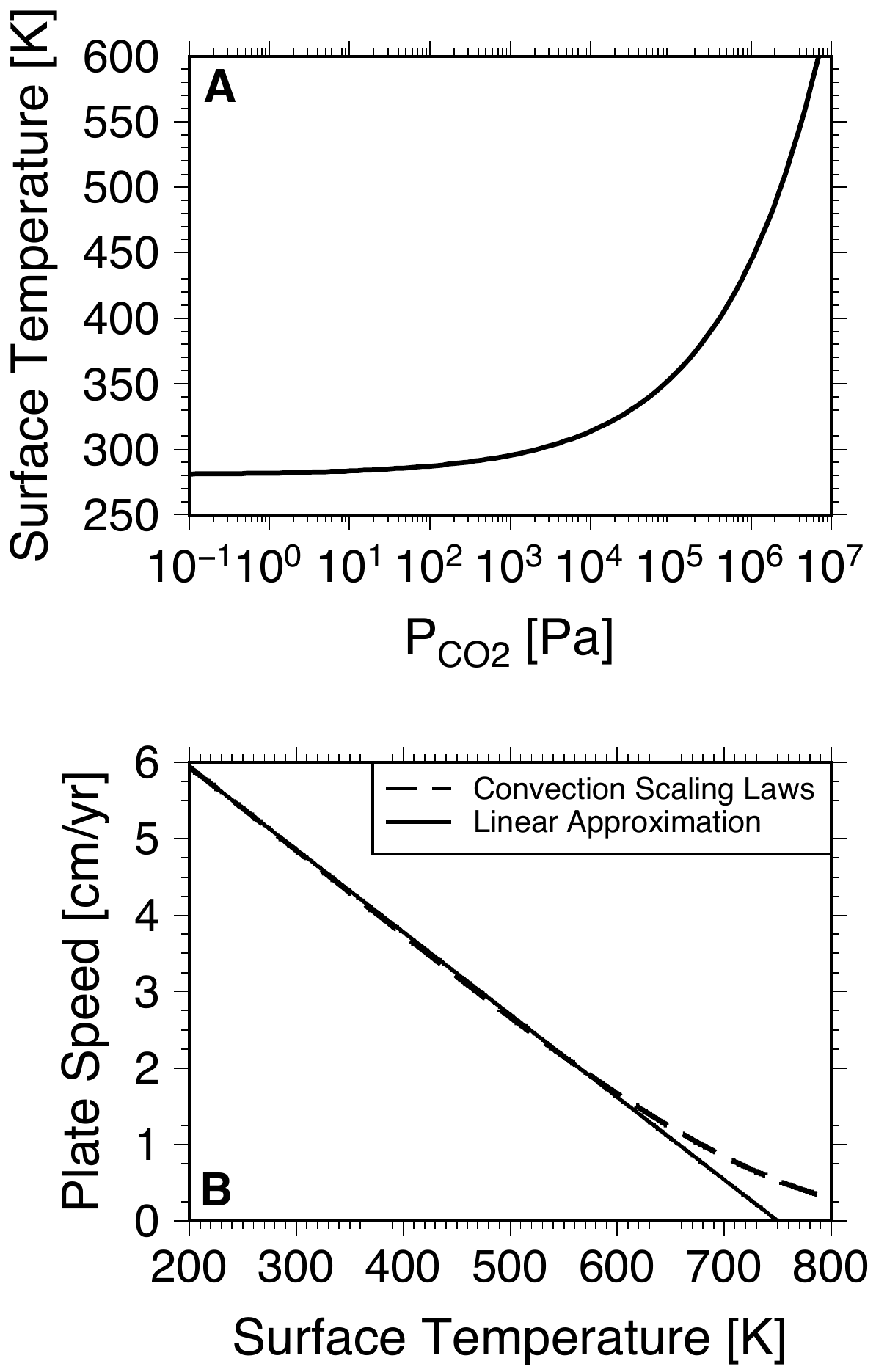}
\caption{\label {fig:v_t_plot} Surface temperature as a function of $P_{CO_2}$ for a present day solar luminosity and albedo (a), and plate speed as a function of surface temperature (b). Both the plate speed parameterization used in this study, equation \eqref{plate_speed}, and the plate speed-surface temperature curve calculated from the scaling laws of \cite{Foley2013_scaling} are shown in (b). }  
\end{figure}

To relate atmospheric CO$_2$ to surface temperature, a simple parameterization from \cite{Walker1981} is used: 
\begin{equation}
\eqlbl{temp}
T = T^* + 2(T_e - T_e^*) + 4.6\left(\frac{P_{CO_2}}{P_{CO_2}^*} \right)^{0.346} - 4.6 
\end{equation}
where $T^* = 285$ K is the present day surface temperature, $T_e$ is the effective temperature, and $T_e^* = 254$ K is the present day effective temperature (Figure \ref{fig:v_t_plot}a). Note that this parameterization also includes the contribution of water vapor to greenhouse warming, assuming that H$_2$O is always saturated in the atmosphere. The effective temperature is related to the absorbed solar radiation as  
\begin{equation}
\eqlbl{te}
T_e = \left(\frac{S(1-A)}{4 \sigma} \right)^{1/4}
\end{equation}
where $S$ is the solar irradiance, $A$ is the albedo, and $\sigma$ is the Stefan-Boltzmann constant. In this study albedo will be held constant for simplicity, so ice-albedo feedbacks, or the currently poorly understood feedbacks involving clouds \citep[e.g.][]{Leconte2013,Wolf2014} are not included. This parameterization provides a good first order approximation to the results of more sophisticated radiative-convective models for an Earth-like planet \citep[e.g.][]{Kasting1986}. As discussed in \S \ref{sec:uncertainties}, using a more advanced climate model may change some of the details of the results, particularly the temperatures calculated at high CO$_2$ levels where equation \eqref{temp} deviates the most from the models of \cite{Kasting1986}, but would not change the overall conclusions of this study. 

Although temperatures reach $\approx 600$ K at $P_{CO_2} \approx 10^7$ Pa, liquid water is still stable at the surface. A true runaway greenhouse, where liquid water cannot exist as a stable phase on the surface, is thought to only be possible through increases in insolation, rather than through the increase in atmospheric opacity brought about by elevated CO$_2$ levels \citep{Nakajima1992}. However, above $\approx 350$ K the atmosphere would be in a moist greenhouse state, where, although liquid water is stable at the surface, high mixing ratios of H$_2$O in the stratosphere lead to rapid water loss to space \citep{Abbot2012}. Such water loss is not modeled in this study; rather I look at the expected climate and tectonic states for terrestrial planets with varying land areas, CO$_2$ budgets, and incoming solar fluxes. However, the implications of the modeled climate states for water loss and habitability are discussed in \S \ref{sec:supply_limit_stab} and \S \ref{sec:discussion}.   

As discussed in \S \ref{sec:intro}, surface temperature can exert a major influence over whether plate tectonics takes place on a planet \citep{Lenardic2008,Landuyt2009a,Foley2012}, and influences the speed of plate tectonics when it is present \citep{Foley2013_scaling,Foley2014_initiation}. To include the effect of climate on plate tectonics, plate speed is assumed to depend on surface temperature using a linear approximation to the scaling laws of \cite{Foley2013_scaling} 
\begin{equation} 
\eqlbl{plate_speed}
v=8.059-0.011T ,
\end{equation}
where high surface temperatures lead to slower plates due to increased grain-growth rates in the lithosphere, which cause effectively stronger plate boundaries. Equation \eqref{plate_speed} gives $v$ in units of cm yr$^{-1}$. Mantle temperature also exerts an influence on plate speed. However, the influence of mantle temperature on plate speed is neglected in this study because it is weaker than the influence of surface temperature \citep{Foley2014_initiation}. Furthermore, ignoring the weaker influence of mantle temperature on plate speed also allows this study to focus solely on the coupling between plate tectonics, carbon cycling, and climate. Figure \ref{fig:v_t_plot}b shows that the plate speed parameterization, equation \eqref{plate_speed}, is an accurate approximation of the full scaling laws from \cite{Foley2013_scaling} up to $\approx 600$ K, corresponding to the highest surface temperatures explored here.  

\section{Results} 
\label{sec:results}

In this section, I first explore the influence the full coupling between plate tectonics and climate has on the development of habitable conditions (\S \ref{sec:stability}), by: testing whether the ability of the long-term carbon cycle to buffer climate against changes in solar insolation is maintained when plate speed depends on surface temperature (\S \ref{sec:buffer}); and by assessing whether habitable climates can develop from arbitrary initial surface temperature and atmospheric CO$_2$ conditions (\S \ref{sec:ic}), or whether different initial conditions prevent temperate climates from forming. Next, I examine the role land area and planetary CO$_2$ budget play in the development and maintenance of habitable climates, and map out the conditions necessary for stable, temperate climates to exist on rocky planets (\S \ref{sec:fland}). Throughout this section, all parameters are held constant (see Tables \ref{tab_param} \& \ref{tab_cons}), save for plate speed which is allowed to vary with surface temperature, so that the coupling between climate and plate tectonics, and the influence of land area and total CO$_2$ budget, can each be studied in isolation. The effect of varying key model parameters is explored in \S \ref{sec:param_test}, and the influence of including both the temperature dependence of seafloor weathering and its supply-limit is tested in \S \ref{sec:sfwt}.   

\subsection{Carbon Cycling and Climate Feedbacks with a Surface Temperature-Dependent Plate Speed}
\label{sec:stability}

\subsubsection{Stability of Climate in Response to Changes in Insolation}
\label{sec:buffer}


\begin{table}
\caption{Table of parameters.}
\label{tab_param}
\begin{threeparttable}
\begin{tabular}{l l l c}
\hline
Symbol & Definition & Baseline Value & Equation \\ 
\hline
$k_c$ & Solubility of CO$_2$ in seawater & $10^7$ Pa (derived) & above \eqref{pco2} \\
$M_{H_2O}$ & Moles of H$_2$O in one ocean mass & $7.6 \times 10^{22}$ mol (derived) & above \eqref{pco2} \\
$L$ & Length of trenches & $6 \times 10^{4}$ km (F05) & \eqref{fsub} \\
$f$ & Fraction of subducted carbon that degasses & 0.5 (A14) & \eqref{fsub} \\
$f_d$ & Fraction of upwelling mantle that degasses & 0.32 (T92) & \eqref{fd} \\
$d_{melt}$ & Depth of melting beneath ridges & 70 km (K06) & \eqref{fd} \\
$F_{sfw}^*$ & Present day seafloor weathering flux & $1.75 \times 10^{18}$ mol Ma$^{-1}$ (M14) & \eqref{fsfw} \\
$\alpha$ & $P_{CO_2}$ exponent for seafloor weathering & $0.25$ (B97) & \eqref{fsfw} \\
$v^*$ & Present day plate speed & 5 cm yr$^{-1}$ (K06) & \eqref{fsfw} \\
$\chi_{cc}$ & Fraction of Mg, Ca, K, and Na in continental crust & $0.08$ (W12) & \eqref{west_2012} \\
$\bar{m}_{cc}$ & Average molar mass of Mg, Ca, K, and Na & 32 g mol$^{-1}$ (derived) & \eqref{convert_west_2012} \\
$E_{max}$ & Maximum erosion rate & $10$ mm yr$^{-1}$ (W13) & \eqref{total_weathering_flux} \\
$\rho_r$ & Regolith density & $2500$ kg m$^{-3}$ (W12) & \eqref{total_weathering_flux} \\
$\beta$ & $P_{CO_2}$ exponent for silicate weathering & 0.55 (D13) & \eqref{my_fw} \\
$a$ & $P_{sat}$ exponent for silicate weathering & 0.3 (D13) & \eqref{my_fw} \\
$E_a$ & Activation energy for silicate weathering & 42 kJ mol$^{-1}$ (B91) & \eqref{my_fw} \\
$F_w^*$ & Present day weathering flux & $12 \times 10^{18}$ mol Ma$^{-1}$ (G99) & \eqref{my_fw} \\
$P_{CO_2}^*$ & Present day atmospheric CO$_2$ & 33 Pa (K84) & \eqref{my_fw} \\
$T^*$ & Present day surface temperature & 285 K (W81) & \eqref{my_fw} \\
$f_{land}^*$ & Present day land fraction & 0.3 (A12) & \eqref{my_fw} \\
$P_{sat0}$ & Reference saturation vapor pressure & 610 Pa (K84) & \eqref{psat} \\
$L_w$ & Latent heat of water & 2469 J g$^{-1}$ (K84) & \eqref{psat} \\
$T_{sat0}$ & Reference temperature & 273 K (K84) & \eqref{psat} \\
$T_e^*$ & Present day effective temperature & 254 K (W81) & \eqref{temp} \\
$A$ & Albedo & 0.31 (derived) & \eqref{te} \\
\hline
\end{tabular}    
\begin{tablenotes}
\item Key for citations: A12=\citep{Abbot2012}, A14=\citep{Ague2014}, B91=\citep{Brady1991}, B97=\citep{Brady1997}, D13=\citep{Driscoll2013}, F05=\citep{Fowler_book}, G99=\citep{Gaillardet1999}, K84=\citep{Kasting1984}, K06=\citep{korenaga2006}, M14=\citep{Mills2014}, T92=\citep{TajikaMatsui1992}, TS=\citep{turc1982}, W81=\citep{Walker1981}, W12=\citep{West2012}, W13=\citep{Willenbring2013}. 
\end{tablenotes}  
\end{threeparttable}   
\end{table}

\begin{table}
\caption{Table of physical constants.}
\label{tab_cons}
\begin{threeparttable}
\begin{tabular}{l l l c}
\hline
Symbol & Definition & Value & Equation \\ 
\hline
$\bar{m}_{CO_2}$ & Molar mass of CO$_2$ & 44 g mol$^{-1}$ (derived) & \eqref{pco2} \\
$g$ & Acceleration due to gravity & 9.8 m s$^{-2}$ (TS) & \eqref{pco2} \\
$V_{mantle}$ & Volume of the mantle & $9.1 \times 10^{20}$ km$^3$ (TS) & \eqref{fd} \\
$A_{Earth}$ & Surface area of Earth & $5.1 \times 10^{14}$ km$^2$ (TS) & \eqref{pco2} \\
$R_g$ & Gas constant & 8.314 J K$^{-1}$ mol$^{-1}$ (N92) & \eqref{my_fw} \\
$\bar{m}_w$ & Molar mass of water & 18 g mol$^{-1}$ (derived) & \eqref{psat} \\
$\sigma$ & Stefan-Boltzmann constant & $5.67 \times 10^{-8}$ W m$^{-2}$ K$^{-4}$ (N92) & \eqref{te} \\
$S^*$ & Solar constant & 1360 W m$^{-2}$ (dP10) & \S \ref{sec:buffer} \\
\hline
\end{tabular}    
\begin{tablenotes}
\item Key for citations: dP10=\citep{dePater2010}, N92=\citep{Nakajima1992}, TS=\citep{turc1982}.  
\end{tablenotes}  
\end{threeparttable}
\end{table}

\begin{figure}
\includegraphics[scale = 0.5]{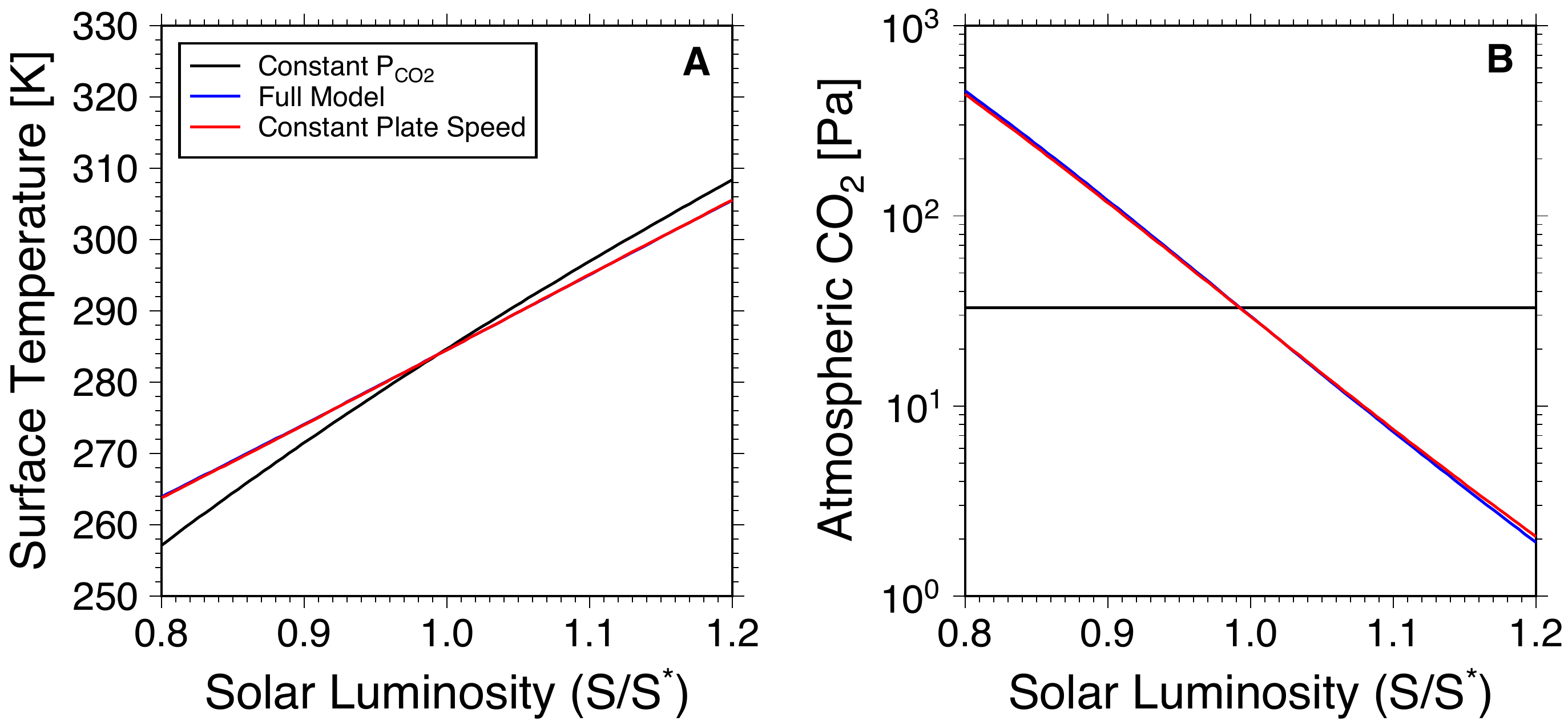}
\caption{\label {fig:buffer} Surface temperature (a) and atmospheric CO$_2$ (b) as functions of insolation normalized by the solar constant. Surface temperature and atmospheric CO$_2$ are determined by solving equations \eqref{rp_eq}-\eqref{raoc_eq} in steady-state for each $S/S^*$. The black line assumes CO$_2$ is constant and equal to the present day value ($P_{CO_2} = P_{CO_2}^*$), the blue line shows the results from the carbon cycle model with a plate speed that depends on surface temperature, and the red line shows the results when plate speed is fixed at 5 cm/yr. Note that the blue and red lines lie almost on top of each other, and are therefore difficult to distinguish. }  
\end{figure}

As outlined in \S \ref{sec:intro}, one of the most important roles the long-term carbon cycle plays in planetary habitability is stabilizing climate against changes in solar luminosity. In this section, the efficacy of the carbon cycle climate buffer, when the influence of surface temperature on plate speed is included, is assessed. The steady-state solution to the global carbon cycle model (equations \eqref{rp_eq}-\eqref{raoc_eq}) as $S$ is varied is shown in Figure \ref{fig:buffer}. Solar luminosities ranging from $0.8S^*$, appropriate for the early Earth \citep{Gough1981}, to $1.2 S^*$, where $S^* = 1360$ Wm$^{-2}$ is the solar constant, are considered. Solving the full model in steady-state assumes that the global carbon cycle reaches steady-state on a timescale shorter than changes in solar luminosity; in reality the two timescales are similar, both on the order of $\sim$ 1 Gyr. A different approach would be to assume that the plate and mantle reservoirs are fixed, and solve for the atmospheric reservoir in steady-state (i.e. setting $F_{sfw} + F_{weather}/2 = fF_{sub} + F_{degas}$). Both approaches were found to give nearly identical results. 

Similar to other studies on climate stabilization and the long-term carbon cycle \citep[e.g.][]{Walker1981}, the temperature dependence of kinetically controlled weathering allows higher atmospheric CO$_2$ levels to build up at low solar luminosity, and lowers atmospheric CO$_2$ levels through higher weathering rates when solar luminosity increases (Figure \ref{fig:buffer}B). As a result, the changes in surface temperature brought about by variations in solar luminosity are less pronounced than the case where atmospheric CO$_2$ is fixed (Figure \ref{fig:buffer}A).  Also consistent with previous studies \citep[e.g.][]{Sleep2001b}, the carbon cycle is not able to produce atmospheric CO$_2$ levels high enough to keep surface temperatures above 273 K at solar luminosities lower than $\approx 0.9S^*$. Other greenhouse gases, such as methane, may be necessary to prevent global glaciation. However, recent results from three-dimensional global climate models have shown that low latitudes can remain ice free at globally averaged surface temperatures of 260 K \citep{Charnay2013,Wolf2013}, and possibly even lower \citep{Abbot2011}, so the ability of the global carbon cycle to maintain liquid water oceans at low solar luminosities may be stronger than previously thought. 
  
Figure \ref{fig:buffer} shows that the dependence of plate speed on surface temperature has almost no influence on the effectiveness of carbon cycle induced climate stabilization. The weathering feedback is by far the dominant factor, and thus the fully coupled climate-plate tectonic system is stable to changes in solar luminosity.  In fact, the influence of surface temperature on plate speed has a very small additional stabilizing effect on long-term climate.  Plate speed increases at cooler temperatures, so lower insolation causes degassing to increase (both at volcanic arcs and mid-ocean ridges), in addition to lowering the weathering rate. A complimentary effect occurs at higher insolation, where the warmer surface temperature slows plates and decreases degassing. However, as plate speed only changes by $\approx 0.2$ cm/yr per 20 degree change in surface temperature, while the weathering flux changes by an order of magnitude, the feedbacks inherent in silicate weathering are by far the dominant factor in long-term climate stabilization.  

\subsubsection{Influence of Initial Conditions}
\label{sec:ic}

The results in \S \ref{sec:buffer} show that habitable climates can be maintained even when plate speed depends on surface temperature. However, they do not demonstrate whether habitable climates develop in the first place. Fully time-dependent evolution models are needed to determine whether different initial conditions, especially different initial atmospheric CO$_2$ concentrations, influence the final surface temperatures reached when the global carbon cycle comes to steady-state. In this section, the full time-dependent form of the carbon cycle model, equations \eqref{rp_eq}-\eqref{raoc_eq}, is solved using three different initial conditions: an initially hot case, where the entire CO$_2$ budget of the planet is initially in the atmosphere and ocean, an initially cold case where the entire CO$_2$ budget initially resides in the mantle, and an intermediary case where half the CO$_2$ starts in the atmosphere and ocean, and half in the mantle. An Earth-like planet is assumed, with a total CO$_2$ budget of $C_{tot} = 2.5 \times 10^{22}$ moles \citep{Sleep2001b}, and a constant land fraction of 0.3. Both the thermal evolution of the mantle and the evolution of solar luminosity are ignored in this section so that the dynamics of the coupled climate-plate tectonic system can be studied in isolation.    

\begin{figure}
\includegraphics[scale = 0.65]{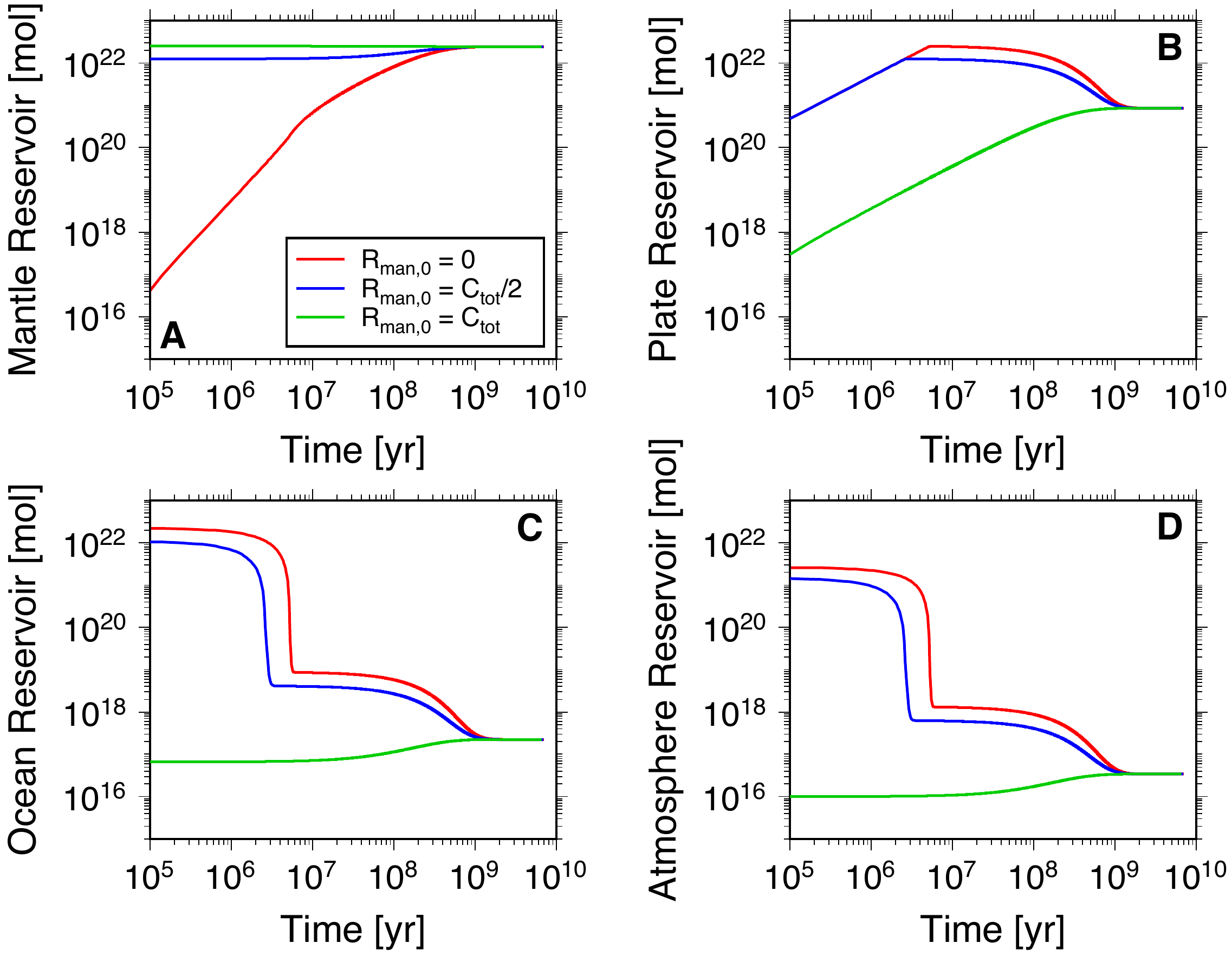}
\caption{\label {fig:earth_model_res} Time evolution of the mantle reservoir (a), plate reservoir (b), ocean reservoir (c), and atmosphere reservoir (d) for an Earth-like model starting from three different initial conditions: all of the carbon in the atmosphere and ocean, i.e. $R_{man,0} = 0$ (red line), half of the carbon in the mantle and half in the atmosphere and ocean, i.e. $R_{man,0} = C_{tot}/2$ (blue line), and all of the carbon in the mantle, i.e. $R_{man,0} = C_{tot}$ (green line). Model parameters are listed in Tables \ref{tab_param} \& \ref{tab_cons}. }  
\end{figure} 

\begin{figure}
\includegraphics[scale = 0.65]{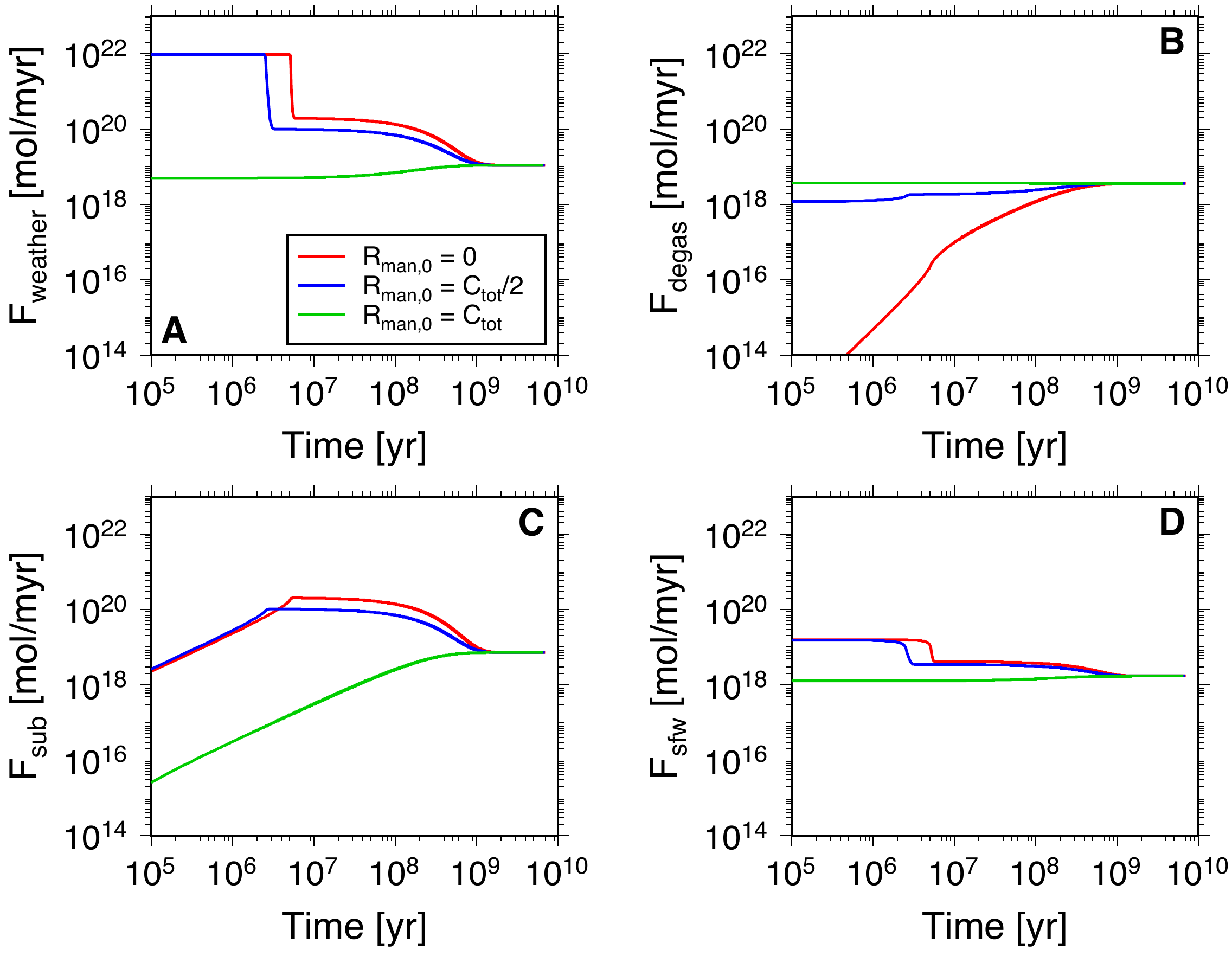}
\caption{\label {fig:earth_model_flux} Time evolution of the terrestrial weathering flux (a), degassing flux (b), subduction flux (c), and seafloor weathering flux (d) for the models shown in Figure \ref{fig:earth_model_res}. The arc flux, $fF_{sub}$, and deep subduction flux, $(1-f)F_{sub}$, are not shown because they are both given by the subduction flux multiplied by a constant. }  
\end{figure} 

\begin{figure}
\includegraphics[scale = 0.5]{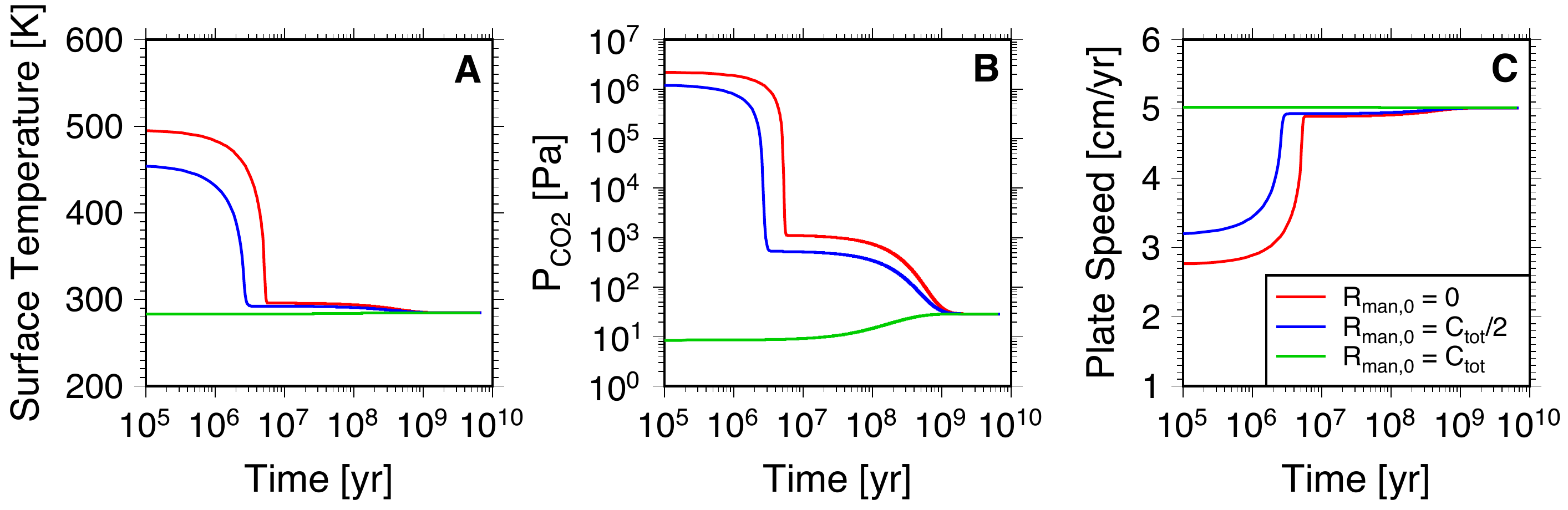}
\caption{\label {fig:earth_model_temp} Time evolution of surface temperature (a), atmospheric CO$_2$ (b), and plate speed (c) for the models shown in Figure \ref{fig:earth_model_res}. }  
\end{figure} 

Figures \ref{fig:earth_model_res}, \ref{fig:earth_model_flux}, and \ref{fig:earth_model_temp} demonstrate that all three initial conditions reach the same steady state after $\sim 1$ Gyr. When atmospheric CO$_2$ is initially high, rapid weathering draws CO$_2$ out of the atmosphere on a 1-10 Myr timescale, dramatically cooling the climate (Figures \ref{fig:earth_model_temp}A \& B).  This rapid CO$_2$ drawdown also ensures that the early moist greenhouse state caused by the high  $P_{CO_2}$ initial condition is short lived, and any water loss during this time would be kept to a minimum (see also \S \ref{sec:supply_limit_stab}). Most of the initial CO$_2$ drawdown takes place via supply-limited weathering (Figure \ref{fig:earth_model_flux}A), as the extremely high initial surface temperatures push the weathering rate into the supply-limited regime. During this initial CO$_2$ drawdown phase, carbon rapidly builds up in the plate reservoir and is gradually subducted into the mantle (seafloor weathering also plays a role, but is much smaller than the continental weathering flux (Figure \ref{fig:earth_model_flux}D)). After the rapid initial draw down of atmospheric CO$_2$, the atmosphere and ocean reservoirs reach a quasi-steady-state as arc degassing is balanced by continental and seafloor weathering. The evolution of carbon in the atmosphere and hydrosphere is then controlled by the gradual loss of carbon from the plate reservoir to the mantle.  As the mantle is regassed by subduction, the degassing flux becomes an important contributor to atmospheric CO$_2$, and the final steady-state is reached as the degassing and arc fluxes balance the terrestrial weathering and seafloor weathering fluxes. Models where half of the CO$_2$ is initially in the mantle and half in the atmosphere and ocean evolve in a similar fashion.  

A different evolution occurs when all of the planet's CO$_2$ initially resides in the mantle. In this case carbon builds up in the atmosphere and ocean through ridge degassing. A quasi-steady-state is quickly reached between mantle degassing and silicate weathering. Weathering also causes carbon to accumulate on the seafloor, and subduct into the mantle. The final steady-state is then reached when the mantle has degassed a sufficient supply of carbon for the deep subduction flux to balance the degassing flux. The timescale for reaching steady-state is thus dictated by the plate speed for all three initial conditions; the plate speed sets how quickly the mantle can be regassed for the hot initial atmosphere models, and how quickly the mantle degasses for the cold initial atmosphere case.  

Despite the different initial conditions leading to different temperature and plate speed evolutions, the same final state is reached. This occurs for two main reasons: 1) the feedbacks inherent in silicate weathering dominate the effect of climate on plate speed; and 2) the influence of climate on plate speed has a stabilizing effect, acting to drive different initial conditions to the same final state. At warm temperatures plate speeds are low, suppressing both mantle degassing and arc volcanism and aiding in the initial drawdown of atmospheric CO$_2$. Likewise when surface temperature is low, a higher plate speed increases the mantle degassing rate, helping to build CO$_2$ back up in the atmosphere.  The effect of variable plate speed is small, however, and the primary reason that different initial conditions do not lead to different final states is the weathering feedback. 

The preceding discussion can be shown more generally by analyzing the steady-state solutions to equations \eqref{rp_eq}-\eqref{raoc_eq}. In particular it can be shown that, for a given total CO$_2$ budget ($C_{tot}$), land fraction, and solar luminosity, only one steady-state solution exists, and that this steady-state solution is stable.  In steady-state, the sum of the four carbon reservoirs must equal the total CO$_2$ budget, $C_{tot} = R_{atm} + R_{oc} + R_p + R_{man}$. Using equations \eqref{rp_eq} and \eqref{rman_eq} in steady-state and Henry's Law, $C_{tot}$ can be written solely as a function of $R_{oc}$, 
\begin{equation}
\begin{split}
\eqlbl{C_tot}
C_{tot} & = R_{oc} \left(1 + \frac{k_c A_{earth}}{(R_{oc} + M_{H_2O})g \bar{m}_{CO_2}} \right) + \\ 
& \left(A_{earth}(1-f_{land}) + \frac{(1-f) V_{man}}{2f_d d_{melt}}\right)(v(R_{oc})L)^{-1} \left [ \frac{F_{weather}(R_{oc})}{2}  + F_{sfw}(R_{oc}) \right ] ,
\end{split}
\end{equation} 
where plate speed, $v$, and both weathering fluxes, $F_{weather}$ and $F_{sfw}$, can be written in terms of the ocean carbon reservoir, rather than the atmospheric carbon reservoir, using Henry's Law.  If the derivative of equation \eqref{C_tot} with respect to $R_{oc}$ is always positive, then there can only be one steady-state solution. 

Taking this derivative term by term, the first term on the right hand side of equation \eqref{C_tot} gives
\begin{equation} 
\frac{\partial (R_{atm} + R_{oc})}{\partial R_{oc}} = 1 + \frac{M_{H_2O}k_c A_{earth}}{(R_{oc} + M_{H_2O} )^2\bar{m}_{CO_2}g}
\end{equation}
which is always positive.  Taking the derivative of the second term on the right hand side of equation \eqref{C_tot} yields 
\begin{equation} 
\eqlbl{Rm_Rp}
\frac{\partial (R_{man} + R_p)}{\partial R_{oc}} = \left(\frac{A_{earth}(1-f_{land})}{L} + \frac{(1-f)V_{man}}{2f_d d_{melt}L}\right) \frac{\partial}{\partial R_{oc}} \left(\frac{F_{weather}}{2v} + \frac{F_{sfw}}{v}  \right),
\end{equation} 
which shows that only the weathering fluxes and plate speed determine the sign of this term; as $f$, $f_{land}$, $f_d$, and $d_{melt}$ are all independent of $R_{oc}$, changing any of these parameters does not lead to multiple steady-states. $F_{weather}$ increases with $R_{oc}$, because a higher ocean carbon reservoir means higher atmospheric CO$_2$ concentrations, warmer temperatures, and a larger weathering flux.  In addition, plate speed decreases with increasing $R_{oc}$ because higher surface temperatures lead to slower plates, so $\partial(F_{weather}/v)/\partial R_{oc}$ is positive due to both the temperature and CO$_2$ dependence of the weathering flux, and the temperature dependence of the plate speed (see Appendix \ref{sec:derivative} for the full derivation of $\partial(F_{weather}/v)/\partial R_{oc}$). Outside of the supply-limited weathering regime, $F_{weather}$ is a much stronger function of $R_{oc}$ than $v$, so the positivity of $\partial(F_{weather}/v)/\partial R_{oc}$ is primarily due to the temperature and $P_{CO_2}$ dependence of the weathering flux. In the supply-limited regime, $\partial F_{weather}/\partial R_{oc} =0$, but the inverse relationship between $R_{oc}$ and $v$ ensures that $\partial(F_{weather}/v)/\partial R_{oc}$ remains positive. Finally, 
\begin{equation} 
\frac{\partial (F_{sfw}/v)}{\partial R_{oc}} = \frac{F_{sfw}^*}{v^* P_{CO_2}^{* \alpha}} \frac{\partial}{\partial R_{oc}} (P_{CO_2}^{\alpha}) = \left(\frac{F_{sfw}^* M_{H_2O}}{v^*} \right) \left(\frac{k_c}{P_{CO_2}^*} \right)^{\alpha} \left(\frac{\alpha R_{oc}^{\alpha-1}}{(M_{H_2O}+R_{oc})^{\alpha+1}} \right) 
\end{equation}
which is also always positive because $\alpha > 0$. Note that $F_{sfw}$ has two different dependencies on $R_{oc}$: increasing $R_{oc}$ increases $P_{CO_2}$ and therefore $F_{sfw}$ directly, but also decreases the plate speed, which acts to slow the seafloor weathering flux. However, as shown by equation \eqref{Rm_Rp}, it is the slope of $F_{sfw}/v$ that determines whether multiple steady-state solutions exist, so the plate-speed effect cancels out. Therefore $\partial C_{tot} / \partial R_{oc}$ is positive, as $F_{weather}/v$ and $F_{sfw}/v$ are both increasing functions of $R_{oc}$, and only one steady-state solution exists. Even with a plate speed dependent on surface temperature, initial conditions do not influence the final state models evolve to (the final steady-state is also shown to be stable in \S \ref{sec:stability_analysis}). An important caveat to this finding is that the carbon cycle model used here implicitly assumes the presence of liquid water oceans and water vapor in the atmosphere. If initial surface temperature and pressure conditions are beyond the critical point for water, it is not clear that silicate weathering, and subsequent CO$_2$ drawdown, can occur; thus, whether carbon cycling can produce habitable climates starting from such high temperature and pressure conditions is unknown. However, as CO$_2$-rock reactions above the liquid water critical point are poorly constrained, such extreme conditions are beyond the scope of this study. 

The only way to produce multiple steady-state solutions with the present model would be for plate speed to have the opposite dependence on surface temperature, where higher temperatures lead to faster plate speeds. In this case both a high surface temperature, high plate speed with rapid degassing, and a low surface temperature, slow plate speed with low degassing steady-state could exist.  However, plate speed would have to be a very strong function of surface temperature in order to counteract the temperature and $P_{CO_2}$ dependence of the weathering fluxes and cause $\partial C_{tot} / \partial R_{oc}$ to change sign. Furthermore, the result that only one steady-state solution exists is robust to uncertainties in the parameters $\beta$, $a$, $E_a$, and $\alpha$, because the sign of all these parameters is well known; in other words the uncertainties in these parameters are not sufficient to introduce hysteresis to the system because they would not change the sign of either $\partial F_{weather} / \partial R_{oc}$ or $\partial F_{sfw} / \partial R_{oc}$. 

\subsection{Transition to Supply-Limited Weathering at Small Land Areas and Large CO$_2$ Inventories}
\label{sec:fland}

The results in \S \ref{sec:stability} show that as long as atmospheric temperature and pressure conditions are below the critical point for liquid water, such that silicate weathering can occur, initial atmospheric CO$_2$ concentrations do not influence the final state reached by an Earth-sized planet with a given exposed land area, total planetary CO$_2$ budget, and solar insolation. The dependence of plate speed on surface temperature does not lead to multiple steady-states or hysteresis in the coupled carbon cycle-plate tectonic system. Furthermore, the dependence of plate speed on surface temperature does not diminish the ability of the carbonate-silicate cycle to buffer surface temperatures against changes in solar luminosity; in fact, climate stabilization is weakly enhanced because degassing is decreased at high surface temperatures. However, the area of exposed land and the planetary CO$_2$ budget can have a major impact on whether hospitable surface temperatures, and the climate buffering effects of the long-term carbon cycle, can be maintained on a planet. 

In particular, if weathering becomes globally supply-limited a planet will not be able to regulate atmospheric CO$_2$ levels through continental weathering; only the weaker seafloor weathering feedback would be active. In this case the planet may enter a hot, moist greenhouse climate where water is lost rapidly to space, either due to high atmospheric CO$_2$ concentrations or the inability of the carbon cycle to drawdown atmospheric CO$_2$ as solar luminosity increases. In this section, I show that at small land fractions and large planetary CO$_2$ inventories, weathering does become supply-limited, leading to planets with hot climates and rapid water loss; with no mechanism for decreasing $P_{CO_2}$ these planets may lose their water and become inhospitable to life. 

\begin{figure}
\includegraphics[scale = 0.6]{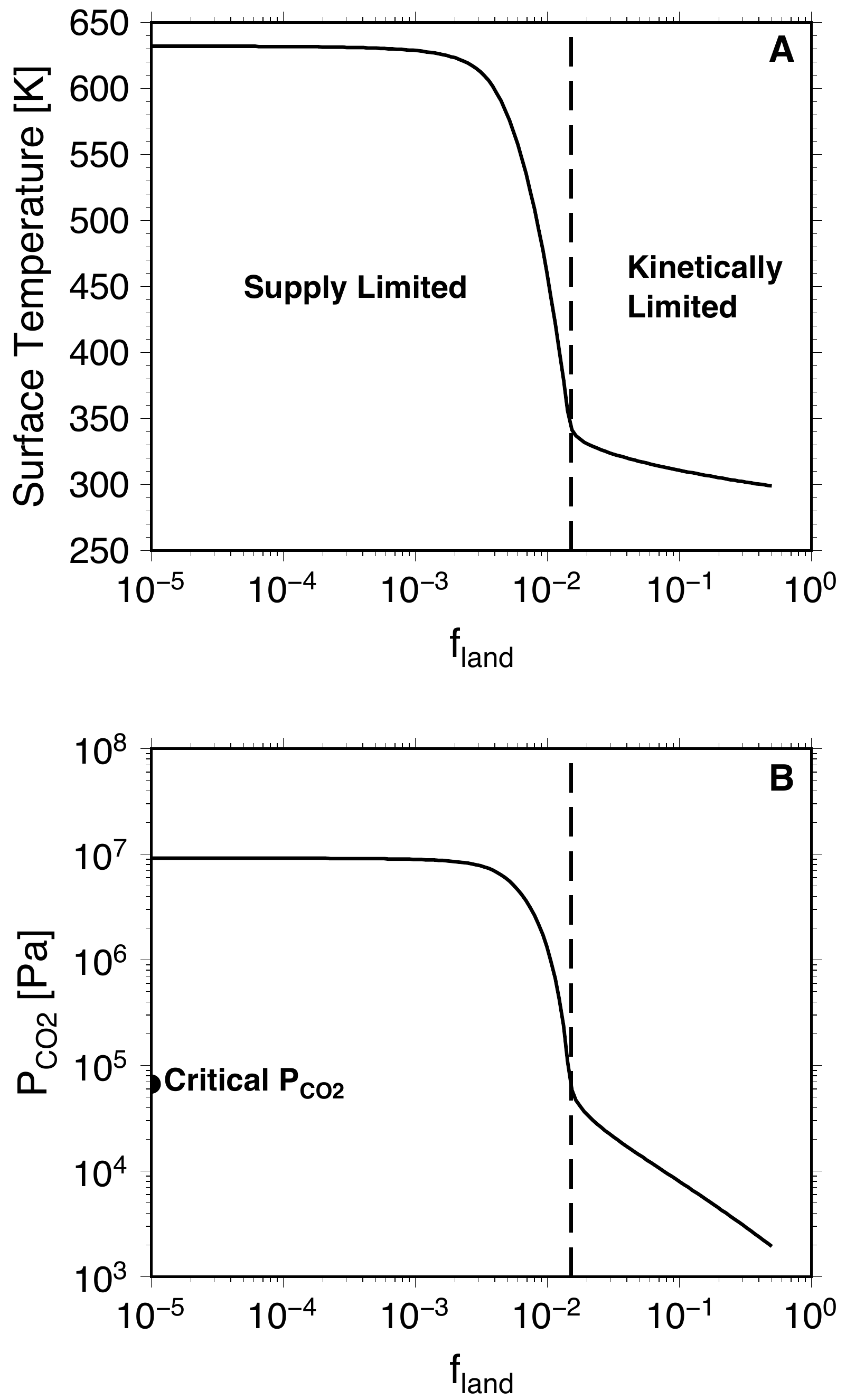}
\caption{\label {fig:supply_lim} Steady-state surface temperature (a) and partial pressure of atmospheric CO$_2$ (b) as a function of land fraction. The boundary between the kinetically-limited and supply-limited weathering regimes is shown with a dashed line, and the critical $P_{CO_2}$, where weathering becomes supply-limited, is shown in panel (b). A total CO$_2$ inventory of $10^{24}$ mol is assumed, and an insolation of $S^*$, the solar constant, is used.  All other model parameters are listed in Tables \ref{tab_param} \& \ref{tab_cons}.}  
\end{figure} 

The transition to a globally supply-limited weathering regime at small land fractions is shown in Figure \ref{fig:supply_lim}. While weathering is kinetically-limited (i.e. for $f_{land} \approx 0.3 - 1.5 \times 10^{-2}$), smaller land fractions lead to a modest increase in both $P_{CO_2}$ and surface temperature. However, once weathering becomes supply-limited, at land fractions lower than $\approx 1.5 \times 10^{-2}$, $P_{CO_2}$ and surface temperature increase sharply as $f_{land}$ decreases, before flattening out as the majority of the planet's total CO$_2$ inventory accumulates in the atmosphere and ocean. The surface temperature and atmospheric CO$_2$ trends, and the transition to supply-limited weathering, can be explained as follows. In the kinetically-limited regime, the change in $P_{CO_2}$ in response to a decrease in $f_{land}$ is governed by 
\begin{equation}
F_{arc} + F_{degas} = \frac{F_w^*}{2} \left(\frac{P_{CO_2}}{P_{CO_2}^*} \right)^{\beta} \exp{\left[\frac{E_a+\bar{m}_w L_w a}{R_g} \left(\frac{1}{T^*} - \frac{1}{T}  \right)\right]} \left(\frac{f_{land}}{f_{land}^*} \right) + F_{sfw}^* \left(\frac{v}{v^*}\right) \left(\frac{P_{CO_2}}{P_{CO_2}^*}\right)^{\alpha} ,
\end{equation}
where the ridge and arc degassing fluxes can be considered constant in this regime. Therefore when $f_{land}$ decreases, $P_{CO_2}$ (and in turn $T$) must increase in order to balance the degassing fluxes. However, as the terrestrial weathering flux is a strong function of $P_{CO_2}$, only a modest increase in atmospheric CO$_2$ is needed to balance the ridge and arc degassing fluxes. The seafloor weathering flux also increases with increasing $P_{CO_2}$, helping to balance the degassing fluxes with only a small increase in atmospheric CO$_2$. However, as seafloor weathering is a weak function of $P_{CO_2}$, it plays a minor role compared to changes in the terrestrial weathering flux.  

The sharp increase in atmospheric CO$_2$ at $f_{land} \approx 1.5 \times 10^{-2}$ represents the transition to globally supply-limited weathering. This transition point occurs at a critical $P_{CO_2}$, where $F_{weather} \approx F_{w_s}$; approximating this condition as $F_{weather} = 0.99 F_{w_s}$, 
\begin{equation}
\eqlbl{crit_co2}
\left(\frac{P_{CO_2}}{P_{CO_2}^*} \right)^{\beta} \exp{\left[\frac{E_a+\bar{m}_w L_w a}{R_g} \left(\frac{1}{T^*} - \frac{1}{T}  \right)\right]} = -\ln{(0.01)} \frac{E_{max} \rho_r \chi_{cc} A_{earth} f_{land}^*}{F_w^* \bar{m}_{cc}}
\end{equation}  
which can be solved numerically to give the critical $P_{CO_2}$ where weathering is supply-limited. Note that the critical $P_{CO_2}$ is not a function of the land fraction, which cancels out in equation \eqref{crit_co2}. When terrestrial silicate weathering is supply-limited, it is fixed at $F_{w_s}$ and therefore no longer a function of atmospheric CO$_2$. As a result, the only way to balance the degassing fluxes is through a higher seafloor weathering flux and by depleting the mantle and plate reservoirs, and hence lowering the CO$_2$ degassing fluxes.  The weak dependence of seafloor weathering on $P_{CO_2}$ means that depleting the mantle and plate reservoirs, thus producing a large increase in atmospheric CO$_2$, is the primary mechanism for bringing the system back into equilibrium. As a result, atmospheric CO$_2$ concentration increases sharply as the land fraction drops below $\approx 1.5 \times 10^{-2}$ (Figure \ref{fig:supply_lim}B). A stronger $P_{CO_2}$ dependence to seafloor weathering, possible if it is directly temperature-dependent, can potentially balance degassing without significant atmospheric CO$_2$ buildup when continental weathering is supply-limited. However, as shown in \S \ref{sec:sfwt}, once seafloor weathering also becomes supply-limited the CO$_2$ drawdown rate will again be unable to balance the degassing rate until the mantle and plate reservoirs are depleted, and hot climates will form. 

\begin{figure}
\includegraphics[scale = 0.6]{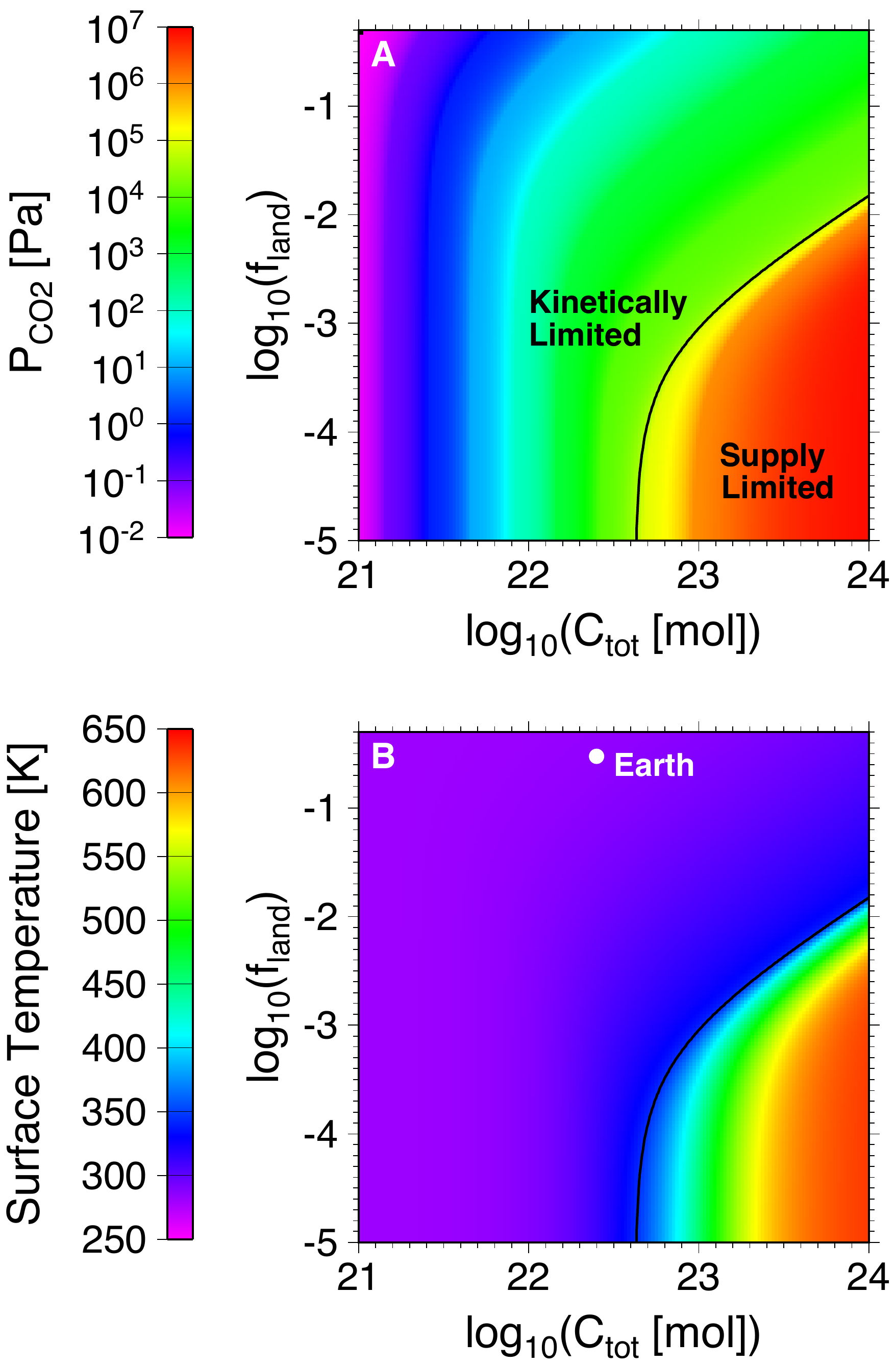}
\caption{\label {fig:supply_lim2} Steady-state partial pressure of atmospheric CO$_2$ (a) and surface temperature (b) as a function of land fraction and total planetary CO$_2$ budget. The boundary between the kinetically-limited and supply-limited weathering regimes is marked by a solid line, and the approximate position of the present day Earth is shown in panel (b). Insolation is assumed equal to the solar constant, and all other model parameters are listed in Tables \ref{tab_param} \& \ref{tab_cons}.}  
\end{figure} 

The transition between globally supply- and kinetically-limited weathering also depends on the planetary CO$_2$ inventory (Figure \ref{fig:supply_lim2}). When $C_{tot}$ decreases, atmospheric CO$_2$ concentrations also decrease in response to lower mantle and arc degassing rates (owing to a smaller amount of CO$_2$ throughout the mantle and surface reservoirs). As a result, the boundary between the kinetically-limited and supply-limited weathering regimes drops to lower $f_{land}$ as $C_{tot}$ decreases (Figure \ref{fig:supply_lim2}A). Below $C_{tot} \approx 5 \times 10^{22}$, the supply-limited regime disappears entirely because seafloor weathering does not allow enough CO$_2$ to build up in the atmosphere for the critical $P_{CO_2}$ to be reached, even for a planet with an infinitesimally small land fraction. Planets with large CO$_2$ inventories and small land fractions therefore lie in the supply-limited weathering regime, resulting in high atmospheric CO$_2$ concentrations and hot climates, while planets with either low CO$_2$ inventories or large land areas, like the Earth, lie in the kinetically-limited weathering regime where the carbon cycle can maintain a temperate climate. Overall, the kinetically-limited regime dominates, at least for the baseline value of $E_{max} = 10$ mm/yr chosen here, and thus the carbon cycle should be able to stabilize climate on most rocky planets.  

\subsubsection{Influence of Erosion Rate and Sensitivity to Model Parameters}
\label{sec:param_test}

\begin{figure}
\includegraphics[scale = 0.5]{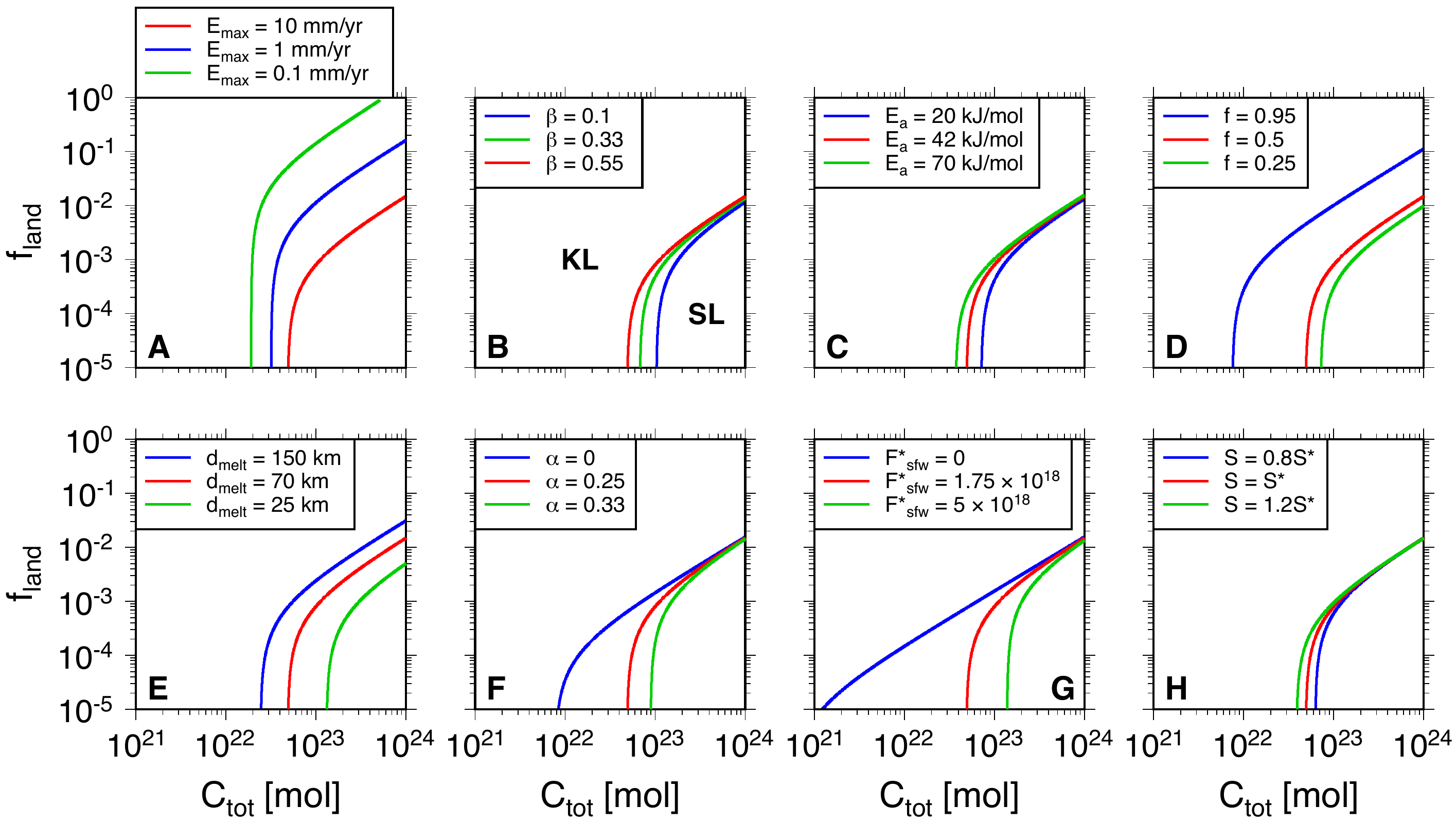}
\caption{\label {fig:param_test} Boundary between the kinetically-limited (KL) and supply-limited (SL) weathering regimes for different $E_{max}$ (a), $\beta$ (b), $E_a$ (c), $f$ (d), $d_{melt}$ (e), $\alpha$ (f), $F_{sfw}^*$ (g), and $S$ (h); for all plots the red curve indicates the standard parameter value used throughout the paper (e.g. in Figures \ref{fig:supply_lim} \& \ref{fig:supply_lim2}).  All model parameters are as specified in Tables \ref{tab_param} \& \ref{tab_cons}, except when explicitly varied. The reference seafloor weathering rate, $F_{sfw}^*$, is given in units of mol/Myr in panel (g). }  
\end{figure}

The boundary between the supply-limited and kinetically-limited weathering regimes is strongly sensitive to the maximum erosion rate ($E_{max}$) used to define the supply limit to weathering, and less sensitive to other model parameters (Figure \ref{fig:param_test}). A lower $E_{max}$ means a lower supply limit to the weathering flux; as a result, the supply-limited weathering regime expands significantly in $C_{tot}-f_{land}$ space. Thus, the ability to sustain high erosion rates is important for effective climate stabilization through the global carbon cycle.  Given that uplift and orogeny are crucial for producing high erosion rates, tectonics may play a key role in keeping planets in the kinetically-limited weathering regime, and therefore enabling the long-term carbon cycle to maintain habitable climates over geologic timescales (see \S \ref{sec:pt_weathering}).

The parameters $\beta$ and $E_a$, which describe the direct dependence of atmospheric CO$_2$ and surface temperature, respectively, on the rate of the weathering reaction, have a relatively weak influence on the boundary between the kinetically-limited and supply-limited regimes (Figures \ref{fig:param_test}B \& C). Lower values of $\beta$, which are applicable to silicate weathering in the presence of land plants \citep[e.g.][]{Berner1994}, slightly decrease the size of the supply-limited regime, because a lower $\beta$ increases the critical CO$_2$ level where weathering becomes supply-limited (see equation \eqref{crit_co2}), therefore allowing kinetically-limited weathering to prevail at smaller land fractions and larger planetary CO$_2$ inventories. Decreasing the activation energy of the weathering reaction, $E_a$, also shrinks the supply-limited regime for the same reason; the critical $P_{CO_2}$ decreases at lower $E_a$. Both larger $f$ and larger $d_{melt}$ expand the supply-limited weathering regime (Figures \ref{fig:param_test}D \& E). When $f$, the fraction of subducted carbon that degasses at arc volcanoes, is larger, atmospheric CO$_2$ levels are higher and the supply limit to weathering is reached at larger land fractions or smaller planetary CO$_2$ inventories. Likewise, a greater $d_{melt}$, the depth of melting at mid-ocean ridges, increases the mantle degassing rate and therefore causes higher atmospheric CO$_2$ concentrations. The influence of $f$ and $d_{melt}$ highlight the role of mantle temperature in maintaining kinetically controlled weathering, as both $f$ and $d_{melt}$ are in principle functions of mantle temperature (where higher mantle temperatures increase both $f$ and $d_{melt}$). 

The parameters governing seafloor weathering, $\alpha$, which describes the dependence on $P_{CO_2}$, and $F_{sfw}^*$, the reference seafloor weathering rate, also have important influences on the weathering regime boundary (Figures \ref{fig:param_test}F \& G). Both smaller $\alpha$ and smaller $F_{sfw}^*$ expand the supply-limited regime, because less efficient seafloor weathering means higher atmospheric CO$_2$ levels, and the critical $P_{CO_2}$ where weathering becomes supply-limited is reached at larger $f_{land}$ or smaller $C_{tot}$. The case for $F_{sfw}^* = 0$ shows where the boundary between the weathering regimes would lie without any seafloor weathering. Finally, varying solar luminosity has only a very minor effect on the weathering regime boundary (Figure \ref{fig:param_test}H); the supply-limited weathering regime is slightly expanded as solar luminosity increases, mainly due to a decrease in the critical $P_{CO_2}$ at higher luminosity. As a result, stellar evolution is not a major factor in determining whether a planet will be able to sustain kinetically-limited weathering. Other parameters, such as $k_c$, the solubility of CO$_2$ in the oceans, and $a$, the exponent describing how changes in temperature influence precipitation rates, were found to have almost no effect on the boundary between the kinetically-limited and supply-limited weathering regimes.  

\subsubsection{Climate Stabilization and Habitability with Supply-Limited Weathering}
\label{sec:supply_limit_stab}

\begin{figure}
\includegraphics[scale = 0.6]{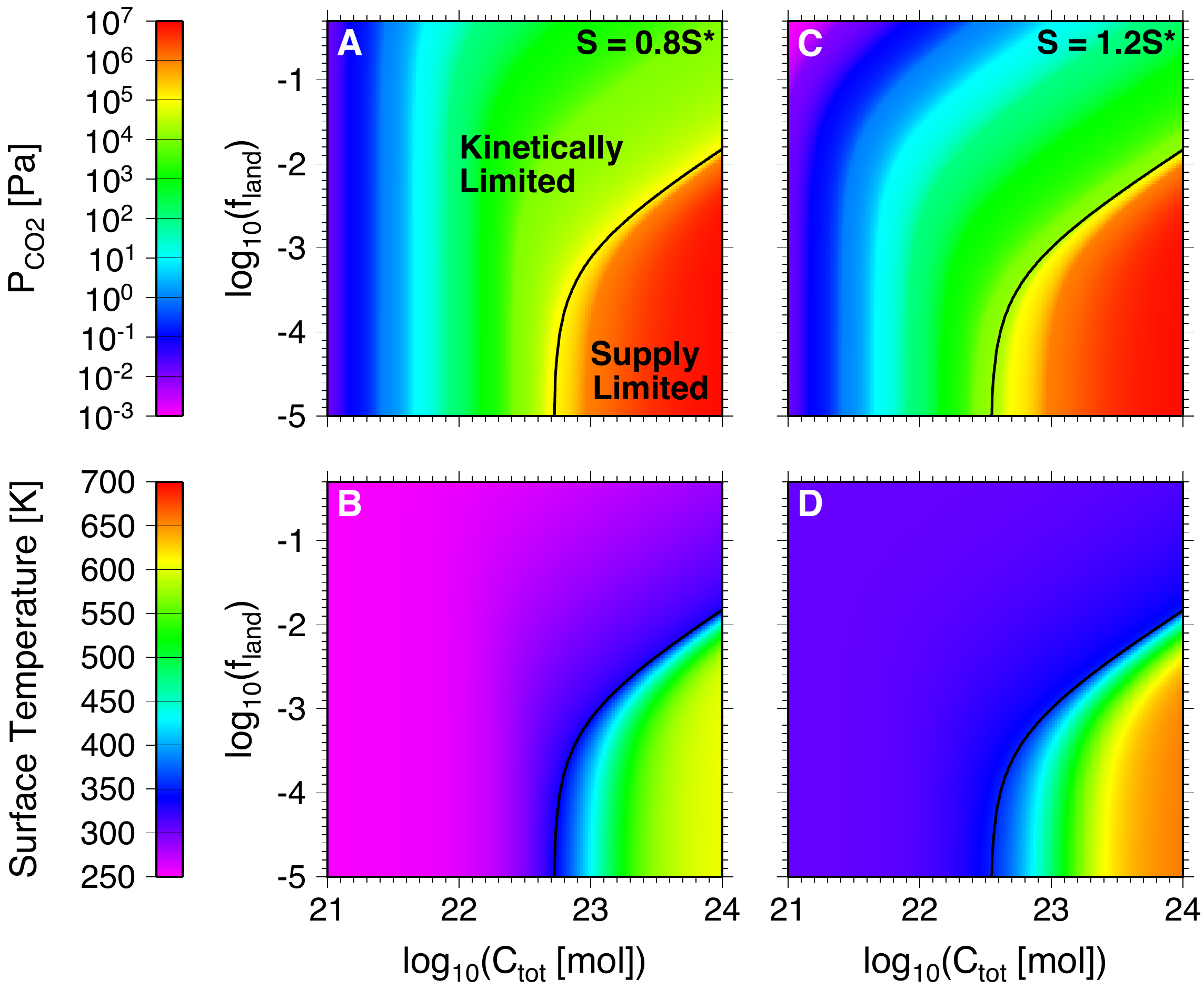}
\caption{\label {fig:co2_S} Steady-state partial pressure of atmospheric CO$_2$ (a) and surface temperature (b) as a function of land fraction and total planetary CO$_2$ budget for $S=0.8S^*$ and $S=1.2S^*$ (panels c \& d). Model parameters are listed in Tables \ref{tab_param} \& \ref{tab_cons}. }  
\end{figure} 

In addition to producing hot climates at the present day solar luminosity, supply-limited weathering also eliminates the carbon cycle's ability to regulate atmospheric CO$_2$ levels in response to changes in luminosity. This is important for planets that lie near the boundary with the kinetically-limited regime, where atmospheric CO$_2$ levels can still be moderate; these planets may have temperate surface temperatures at low solar luminosities, but as luminosity increases, the inability to draw down atmospheric CO$_2$ will cause temperatures to rise significantly. Likewise, planets within the supply-limited regime that sit near the inner edge of the habitable zone will experience much higher surface temperatures than those near the outer edge. Climate stabilization in the kinetically- and supply-limited weathering regimes is shown in Figure \ref{fig:co2_S}. Within the kinetically-limited weathering regime, increasing solar luminosity is counteracted by a decrease in $P_{CO_2}$, thanks to the weathering feedback (Figures \ref{fig:co2_S}A \& C). As a result, surface temperatures remain moderate (Figures \ref{fig:co2_S}B \& D). However in the supply-limited regime, atmospheric CO$_2$ cannot adjust to changes in solar luminosity, and temperatures increase more sharply as luminosity increases. 

\begin{figure}
\includegraphics[scale = 0.75]{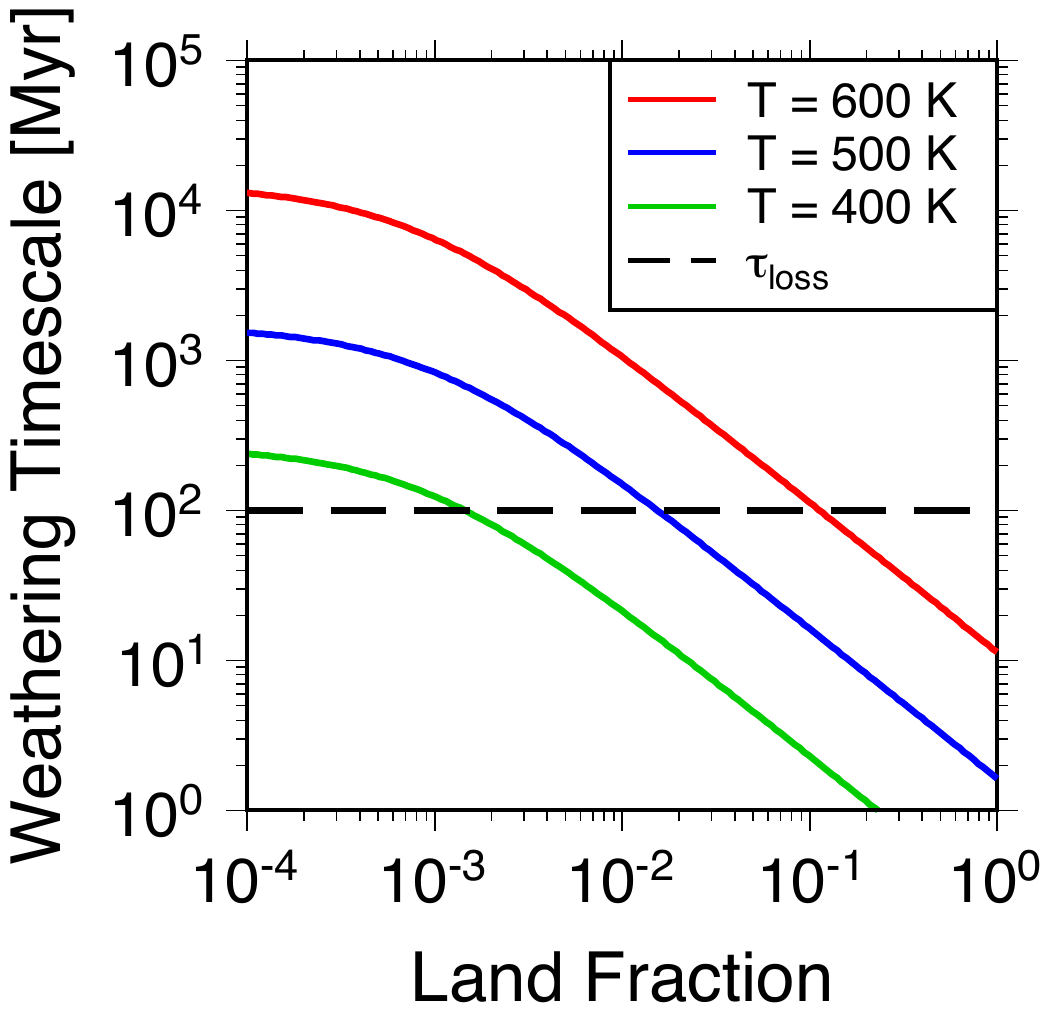}
\caption{\label {fig:tau_w} Weathering timescale, equation \eqref{tau_w}, as a function of land fraction for three different surface temperatures with a present day solar luminosity: $T = 400$ K (green line), $T = 500$ K (blue line), and $T=600$ K (red line).  These surface temperatures correspond to atmospheric CO$_2$ levels of $P_{CO_2} \approx 4$ bar, $P_{CO_2} \approx 23.5$ bar, and $P_{CO_2} \approx 70$ bar, respectively. The 100 Myr timescale for water loss, $\tau_{loss}$, is shown as a dashed line. }  
\end{figure} 

The hot, CO$_2$ rich atmospheres that result from supply-limited weathering have negative implications for planetary habitability. First, the typical surface temperatures estimated for the supply-limited regime, 400-600 K, are higher than any known life form on the modern Earth can tolerate \citep{Takai2008}, and thus life may not be possible at these conditions. Second, although these surface temperatures are within the liquid water stability field, they would place planets in a moist greenhouse climate state where photo dissociation of H$_2$O and hydrogen escape leads to rapid water loss \citep[e.g.][]{Kasting1988}. As a result, planets in the supply-limited weathering regime could lose their oceans, rendering them uninhabitable. However, whether complete water loss will occur in practice is unclear because as sea level drops more land will be exposed, enhancing continental weathering and potentially drawing enough CO$_2$ out of the atmosphere to stop water loss. If planets that originally lie in the supply-limited weathering regime can reestablish temperate, stable climates with higher land fractions before their oceans are completely lost to space, they will once again be habitable (i.e. the waterworld self-arrest mechanism proposed by \cite{Abbot2012}). 

A detailed model of water loss is beyond the scope of this study, so the likelihood of waterworld self-arrest on supply-limited planets is assessed using a simple order of magnitude estimate. Specifically, the timescale for complete water loss is compared to the timescale for CO$_2$ drawdown via silicate weathering. If Earth were completely covered in oceans, the timescale for water loss is estimated at $\sim 100$ Myr \citep{Abbot2012}. To determine the timescale for CO$_2$ drawdown, I assume a planet that initially lies within the supply-limited regime, and therefore has a hot, CO$_2$ rich atmosphere, undergoes water loss to a given land fraction. The weathering timescale, $\tau_w$, is therefore  \begin{equation} 
\eqlbl{tau_w}
\tau_w = \frac{R_{atm}(T) + R_{oc}(T)}{(1/2)F_{w_s}(f_{land}) + F_{sfw}(T)} . 
\end{equation} 
Carbon dioxide dissolved in the oceans is included because rapid equilibration between the atmosphere and ocean means that CO$_2$ from both reservoirs must be drawn down to significantly cool climate. Furthermore, continental weathering is supply-limited for all land fractions, and thus the CO$_2$ drawdown rate is primarily controlled by the land fraction, because the planets being considered here have CO$_2$ rich atmospheres. As shown by equation \eqref{crit_co2}, the transition to supply-limited weathering is determined solely by $P_{CO_2}$, and is independent of $f_{land}$. Weathering can only decrease from the supply-limit as atmospheric CO$_2$ is drawn down, so using the supply-limit to continental weathering in equation \eqref{tau_w} gives the most optimistic estimate. 

For a planet with $T=600$ K, $\tau_w$ does not reach 100 Myr until the land fraction is $\approx 0.1$ (Figure \ref{fig:tau_w}). At this point significant water loss will have occurred, and the remaining oceans may be lost before weathering can cool the climate (i.e. the water loss timescale at this point will be less than 100 Myrs, because most of the oceans have already been lost). The atmospheric and oceanic CO$_2$ reservoirs will also have declined by the time $f_{land}$ reaches $\approx 0.1$, but given the long weathering timescales this effect is likely negligible. However, with $T=400$ K, corresponding to a planet that originally sat near the boundary between supply- and kinetically-limited weathering, the weathering timescale drops below 100 Myr at land fractions of $\sim 10^{-3}$, and waterworld self-arrest appears to be quite likely. Thus the amount of CO$_2$ that weathering must draw down, which correlates to how far into the supply-limited regime a planet lies, is crucial to determining whether water loss can stabilize climate before the oceans are lost. A planet's bathymetry will also be important. If seafloor topography is dominated by a few tall islands, with little remaining topography, waterworld self-arrest will be less effective because nearly all of the planet's water will be lost before a significant portion of land is exposed. However, if there are large submerged continents surrounded by deep basins, a significant supply of water will remain even after a large land fraction has been exposed. Waterworld self-arrest may be viable, especially for planets that originally sit near the boundary with kinetically-limited weathering, but more sophisticated models taking into account bathymetry are needed to place better constraints on this process. 
 
\subsubsection{Temperature-dependent Seafloor Weathering}
\label{sec:sfwt}

In this section I show that including temperature-dependent seafloor weathering and an upper limit on the basalt CO$_2$ reservoir does not significantly impact the transition to supply-limited weathering, or the predicted steady-state climates in $f_{land}$ - $C_{tot}$ space, justifying the simpler model used in the rest of this study where seafloor weathering depends solely on $P_{CO_2}$. 
The upper bound on the basalt CO$_2$ reservoir, $R_{bas_{max}}$, is given by calculating the number of moles of CO$_2$ that are stored by complete carbonation of the seafloor to a certain depth: 
\begin{equation}
\eqlbl{bas_max}
{R_{bas_{max}}}= \frac{d_{carb} A_{Earth} (1-f_{land}) \rho \chi_{bas}}{\bar{m}_{bas}}
\end{equation}
where $d_{carb}$ is the depth of complete basalt carbonation, $\rho$ is the density of basalt, $\chi_{bas}$ is the fraction of reactable elements in basalt (CaO, MgO, and FeO), and $\bar{m}_{bas}$ is the average molar mass of CaO, MgO, and FeO \citep{Sleep2001a}. From \cite{Sleep2001a}, $\chi_{bas} \approx 0.3$, $\bar{m}_{bas} \approx 55$ g mol$^{-1}$, and a baseline value of $d_{carb} \approx 500$ m is set based on the depth that carbonation extends to on the modern day Earth \citep{Alt1999}.  With a negligible land fraction $R_{bas_{max}} \approx 4 \times 10^{21}$ mol (see Table \ref{tab_sfwt}). 

\begin{table}
\caption{Table of variables and parameters for temperature-dependent seafloor weathering.}
\label{tab_sfwt}
\begin{threeparttable}
\begin{tabular}{l l l c}
\hline
Symbol & Definition & Baseline Value & Equation \\ 
\hline
$d_{carb}$ & Depth of complete basalt carbonation & 500 m (S01) & \eqref{bas_max} \\
$\rho$ & Basalt density & $2800$ kg m$^{-3}$ (TS)  & \eqref{bas_max} \\
$\chi_{bas}$ & Fraction of CaO, MgO, and FeO in basalt & 0.3 (S01) & \eqref{bas_max} \\
$\bar{m}_{bas}$ & Molar mass of CaO, MgO, and FeO & 55 g mol$^{-1}$ (derived) & \eqref{bas_max} \\
$R_{bas_{max}}$ & Maximum size of basalt reservoir & $4 \times 10^{21}$ mol (S01) & \eqref{bas_max} \\
$R_{sed}$ & Seafloor sedimentary carbon reservoir & - & \eqref{rsed_eq_new} \\
$R_{bas}$ & Seafloor basalt carbon reservoir & - & \eqref{rbas_eq_new} \\
$F_{sfw_T}$ & Temperature-dependent seafloor weathering flux & - & \eqref{fsfw_T1} \\
$E_{sf}$ & Activation energy for seafloor weathering & 50 kJ mol$^{-1}$ (B97) & \eqref{fsfw_T1} \\
\hline
\end{tabular}    
\begin{tablenotes}
\item Key for citations: B97=\citep{Brady1997}, S01=\citep{Sleep2001a}.  
\end{tablenotes}  
\end{threeparttable}

\end{table}

In order to incorporate temperature-dependent seafloor weathering with a supply-limit, as determined by equation \eqref{bas_max}, the carbon cycle model (equations \eqref{rp_eq}-\eqref{raoc_eq}) is modified as follows: the plate reservoir is separated into a basalt reservoir, $R_{bas}$, and a sedimentary reservoir, $R_{sed}$. Seafloor weathering is assumed to supply carbon solely to the basalt reservoir, and terrestrial weathering supplies carbon solely to the sedimentary reservoir.  The modified carbon cycle equations are therefore 
\begin{equation}
\eqlbl{rsed_eq_new}
\frac{\mathrm{d} R_{sed}}{\mathrm{d}t} = \frac{F_{weather}}{2} - \frac{R_{sed}vL}{A_{Earth}(1-f_{land})}
\end{equation}
\begin{equation}
\eqlbl{rbas_eq_new}
\frac{\mathrm{d} R_{bas}}{\mathrm{d}t} = {F_{sfw_{T}}} - \frac{R_{bas}vL}{A_{Earth}(1-f_{land})}
\end{equation}
\begin{equation}
\eqlbl{rman_eq_new}
\frac{\mathrm{d} R_{man}}{\mathrm{d}t} = \frac{(1-f)vL(R_{bas}+R_{sed})}{A_{Earth}(1-f_{land})} - F_{degas}
\end{equation}
\begin{equation}
\eqlbl{raoc_eq_new}
\frac{\mathrm{d} (R_{atm} + R_{oc}) }{\mathrm{d}t} = \frac{fvL(R_{bas}+R_{sed})}{A_{Earth}(1-f_{land})} + F_{degas} - \frac{F_{weather}}{2} - {F_{sfw_{T}}} , 
\end{equation} 
where 
\begin{equation}
\eqlbl{fsfw_T1}
F_{sfw_T} = F_{sfw}^* \left(\frac{v}{v^*}\right) \left(\frac{P_{CO_2}}{P_{CO_2}^*}\right)^{\alpha} \exp{\left(\frac{E_{sf}}{R_g} \left(\frac{1}{T^*} - \frac{1}{T}  \right)\right)}  
\end{equation}
for $R_{bas} < R_{bas_{max}}$ and 
\begin{equation}
\eqlbl{fsfw_T2}
F_{sfw_T} = \frac{R_{bas}vL}{A_{Earth}(1-f_{land})}
\end{equation}
for $R_{bas} = R_{bas_{max}}$. A seafloor weathering activation energy of $E_{sf} = 50$ kJ mol$^{-1}$ is assumed, in the middle of the range of activation energies given by \cite{Brady1997}. 

\begin{figure}
\includegraphics[scale = 0.6]{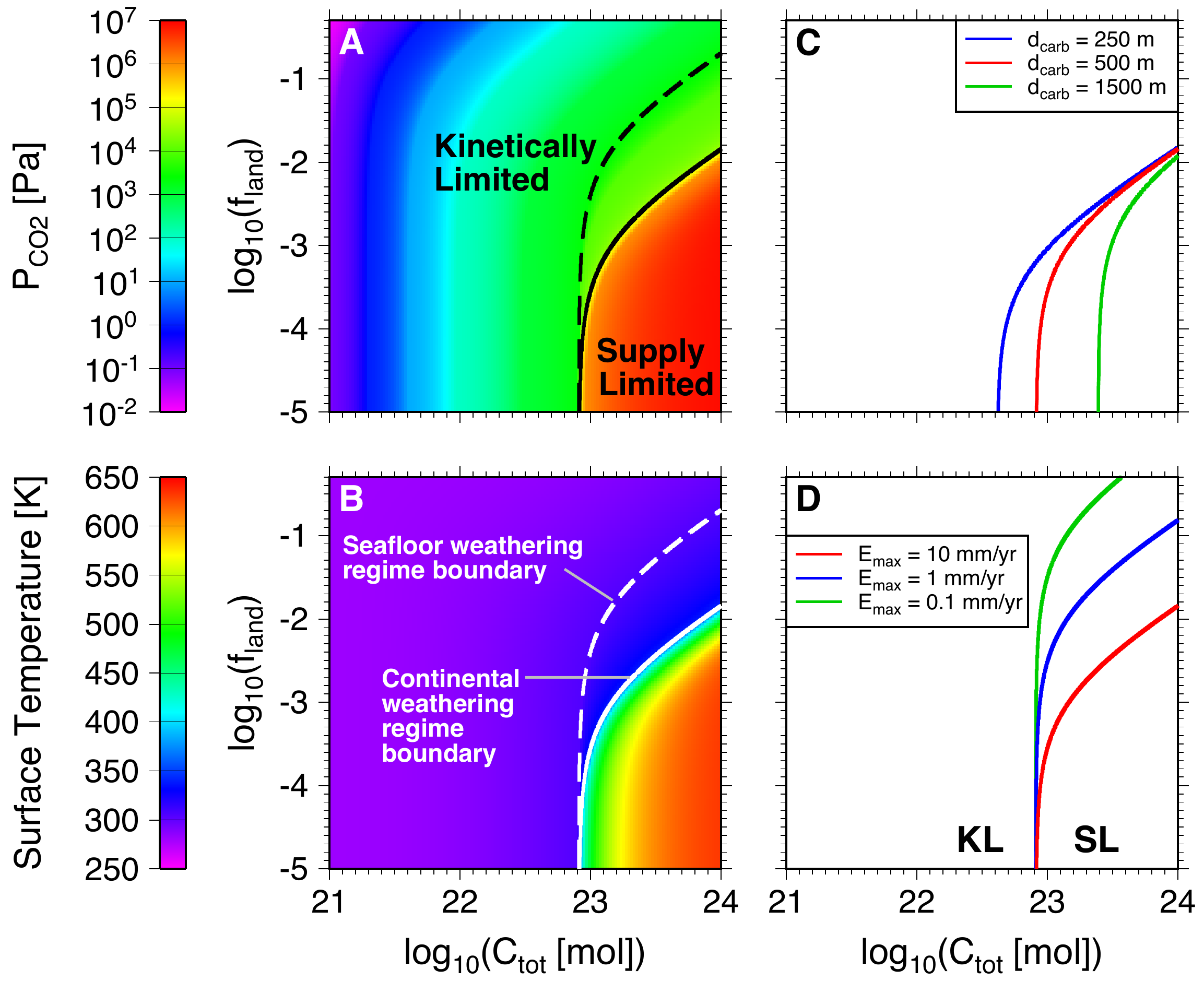}
\caption{\label {fig:sfwt} Steady-state partial pressure of atmospheric CO$_2$ (a), and surface temperature (b) for the modified carbon cycle model including temperature-dependent seafloor weathering (equations \eqref{rsed_eq_new}-\eqref{fsfw_T2}; see Table \ref{tab_sfwt} for parameter values). The dashed line shows when seafloor weathering becomes supply-limited (i.e. when $R_{bas} = R_{bas_{max}}$), and the solid line shows when continental weathering becomes supply limited. Weathering is considered to be globally supply-limited when both seafloor and terrestrial weathering are supply-limited. The boundary between globally supply-limited (SL) and kinetically-limited (KL) weathering is plotted for varying $d_{carb}$ (c) and $E_{max}$ (d). }
\end{figure}

Solving for the steady-state solutions to equations \eqref{rsed_eq_new}-\eqref{raoc_eq_new} as a function of $f_{land}$ and $C_{tot}$ shows that adding a direct temperature feedback and a limit to the size of the basalt reservoir does not significantly change the predicted climate states (Figure \ref{fig:sfwt}A,B). Supply-limited weathering still leads to extremely hot climates at low land fractions and high CO$_2$ budgets, because the basalt reservoir is not large enough to prevent the accumulation of a CO$_2$ rich atmosphere at these conditions. The supply-limited regime is slightly smaller when compared to the models without temperature-dependent seafloor weathering (Figure \ref{fig:supply_lim2}), with higher total CO$_2$ budgets required for weathering to become supply-limited.  The difference is minor because even without a direct temperature feedback, seafloor weathering still acts as a major carbon sink at low land fractions and large CO$_2$ inventories. The only way to shrink the supply-limited regime further through seafloor weathering is with a larger carbonation depth, as shown by Figure \ref{fig:sfwt}C (increasing either $E_{sf}$ or $\alpha$ does not effect the weathering regime boundary because the boundary is dictated by the maximum size of the basalt CO$_2$ reservoir). However, $d_{carb} > 6$ km would be required to completely eliminate the supply-limited weathering regime for the range of $C_{tot}$ and $f_{land}$ explored here. Complete carbonation of basalt to over 6 km depth, or even beyond the top few hundred meters, is not likely because hydrothermal carbonate formation is generally confined to the near surface layer of pillow basalts where permeability is high and significant hydrothermal circulation can occur \citep{Alt1999}. High surface temperatures and atmospheric CO$_2$ concentrations will make the accessible basalt more reactive, but are unlikely to cause deeper hydrothermal circulation because the seafloor permeability structure would be unaffected. Moreover, the supply-limited weathering regime is still strongly influenced by the maximum erosion rate, with lower erosion rates leading to a larger supply-limited regime, even for the modified seafloor weathering flux used in this section (Figure \ref{fig:sfwt}D).

\section{Discussion} 
\label{sec:discussion}

\subsection{Implications for Planetary Habitability} 

Overall, the results of this study have generally positive implications for planetary habitability. The results show that the inverse dependence of plate speed on surface temperature proposed by \cite{Foley2013_scaling} and \cite{Foley2014_initiation} does not inhibit the carbon cycle's ability to regulate climate (\S \ref{sec:buffer}). Furthermore, the results also show that as long as silicate weathering can occur, initial atmospheric CO$_2$ concentrations, and therefore temperatures, do not influence the final steady-state reached (\S \ref{sec:ic}); planets with a given exposed land area, total CO$_2$ inventory, and receiving a given solar irradiance will evolve to the same surface temperature and $P_{CO_2}$ regardless of the initial distribution of carbon between mantle and surface reservoirs. Thus, an initially hot climate is not an obstacle for developing temperate surface temperatures and active plate tectonics on a planet. Other factors, however, can be an obstacle for the development of a habitable climate. Planets with small land fractions and large CO$_2$ inventories are found to enter a supply-limited weathering regime, where weathering depends only on the erosion rate and exposed land area, and not on surface temperature and atmospheric CO$_2$ (\S \ref{sec:fland}). As a result, planets in the supply-limited weathering regime tend to develop hot, CO$_2$ rich climates because the carbonate-silicate cycle can no longer regulate atmospheric CO$_2$ levels. Such hot climates are less favorable for life and could lead to significant water loss due to photodissociation of H$_2$O in the upper atmosphere and hydrogen escape to space. Furthermore, carbon cycling in the supply-limited weathering regime can no longer regulate atmospheric CO$_2$ levels in response to changes in solar luminosity (\S \ref{sec:supply_limit_stab}). 

Nevertheless, the supply-limited regime only takes up a relatively small portion of the $f_{land}$-$C_{tot}$ parameter space. Thus, only planets that have significantly larger amounts of both water and carbon dioxide than the Earth would be at risk of losing the carbon cycle's stabilizing climate feedbacks due to supply-limited weathering. As long as erosion rates are high, planets with a CO$_2$ budget similar to Earth's can develop habitable climates, with kinetically-limited weathering, at land fractions as small as $10^{-4}-10^{-3}$. As a result, only a handful of Hawaiian type island chains would be needed to maintain an active carbon cycle with stabilizing climate feedbacks (the Hawaiian islands alone have a land fraction of $\approx 5 \times 10^{-5}$). A significant percentage of Earth-like planets within the habitable zone should therefore be able to develop temperate climates and plate tectonics, and maintain these habitable surface temperatures over geologic timescales. Moreover, some planets that fall into the supply-limited weathering regime may even be able to reestablish stable, temperate climates as water loss exposes more land (\S \ref{sec:supply_limit_stab}). In particular, planets that originally sit near the boundary with kinetically-limited weathering only need to expose modest amounts of land before the timescale to cool the climate through weathering becomes shorter than the timescale for complete water loss. This waterworld self-arrest mechanism may not work for all planets, however. Planets that initially lie deep within the supply-limited regime have much higher atmospheric CO$_2$ concentrations, resulting in much longer climate cooling timescales; thus complete water loss may occur before silicate weathering can significantly cool the climate on these planets. Even so, the possibility of waterworld self-arrest further increases the odds that a significant fraction of planets forming within the habitable zone will indeed be habitable over a large portion of their geologic lifetimes.  

\subsubsection{Influence of Planet Size}
\label{sec:size}

Although this study focuses solely on Earth-sized planets, some preliminary conclusions about rocky planets of different sizes can be drawn. Planet size influences whether plate tectonics can occur on a planet \citep[e.g.][]{ONeill2007c,Foley2012}, the structure of the atmosphere and hence the strength of the radiative forcing provided by greenhouses gasses such as CO$_2$ and H$_2$O \citep[e.g.][]{Kasting1993,Kopp2014}, and the rate at which water is lost to space \citep[e.g.][]{Hunten1973,Melosh1989}. The influence of size on the propensity for plate tectonics on a planet is controversial \citep{Valencia2007b,ONeill2007,Kite2009,Valencia2009,Korenaga2010a,vanheck2011,Foley2012,Stein2013,Noack2013}, because the mechanism that leads to plate tectonics on Earth is unknown \citep[e.g.][]{Tackley2000,berco2003}, and because key material properties, such as viscosity and thermal conductivity, are not well constrained at the temperature and pressure conditions expected for super-Earth planets \citep{Karato2010,Stamenkovic2011}.  

With the grain-damage mechanism for generating plate tectonics from mantle convection, \cite{Foley2012} found that larger planets are more likely to have plate tectonics because larger planets have more vigorously convecting mantles and therefore larger driving forces for grainsize reduction. As a result, small planets, like Mars, may not able to develop plate tectonics even with cool surface temperatures. Furthermore, small planets lose their water to space more easily because of their smaller gravity \citep[e.g.][]{Kasting2003}. Thus small rocky planets are unlikely to maintain liquid water oceans because of both a lack of plate tectonics and an inability to prevent rapid water loss to space. On the other hand larger planets, in addition to being more likely to have plate tectonics, also have a larger habitable zone, because higher planet mass reduces the greenhouse effect of H$_2$O, thereby moving the inner edge of the habitable zone to lower orbital distances \citep{Kopp2014}. However, the volatile content of super-Earths will also be important; if these planets are volatile rich they could end up in a supply-limited weathering regime, due to a small land area and a high planetary CO$_2$ budget. Furthermore, the conclusion from \cite{Foley2012} that plate tectonics is more likely on larger planets should be considered preliminary until better constraints are placed on the material properties of super-Earth mantles, and these constraints are included in numerical models of mantle convection with grain-damage.  

\subsection{Model Uncertainties}
\label{sec:uncertainties}

\subsubsection{Parameter Uncertainties}

The carbon cycle model used in this study relies on a large number of parameters (Table \ref{tab_param}), many of which are not well constrained. However, as seen in \S \ref{sec:results}, the overall results of this study hold for a wide range of model parameters, and are therefore robust. The result from \S \ref{sec:stability}, that the dependence of plate speed on surface temperature does not significantly alter the negative climate feedbacks of the carbon cycle, nor does it add hysteresis to the system, holds for a wide parameter range. The reason the dependence of plate speed on surface temperature has such a minor impact on the long-term carbon cycle's climate buffering capacity is that the influence of temperature on the weathering flux is significantly stronger than its influence on plate speed. Either the activation energy for the weathering reaction ($E_a$) would have to be significantly lower than what is inferred from laboratory results and field studies, or the dependence of plate speed on surface temperature would have to be far stronger than what is inferred from numerical convection models, for variations in plate speed to significantly alter the carbon cycle. Even then, the influence of surface temperature on plate tectonics has a stabilizing effect, because higher surface temperatures lead to slower plates, and therefore lower degassing rates. Thus, some form of climate stabilization would be preserved as long as plate speed depends inversely on surface temperature, as expected by mantle convection studies with grain-damage \citep{Foley2013_scaling,Foley2014_initiation}.     

The result of \S \ref{sec:ic}, that there is only one steady-state for a planet with a given land fraction and total CO$_2$ budget to evolve to, is similarly robust. This result is independent of model parameters, as long as the two most fundamental aspects of the model, that the weathering flux never decreases with increasing atmospheric CO$_2$ and that the plate speed decreases with increasing surface temperature, hold true. Note that the results of \S \ref{sec:ic} still hold even when temperature-dependent seafloor weathering and a limit to basalt carbonation are included because the quotient of the seafloor weathering flux and the plate velocity would still never be a decreasing function of $R_{oc}$. The only plausible way to introduce hysteresis into the coupled carbon cycle-plate tectonic system would be through additional feedbacks, not included in the present model; furthermore any new feedbacks would have to be very strong in order to counteract the weathering feedbacks. Finally, the position of the boundary between the kinetically-limited weathering and supply-limited weathering regimes changes with key model parameters (\S \ref{sec:param_test}) or with the inclusion of temperature-dependent seafloor weathering (\S \ref{sec:sfwt}), but the overall trend of supply-limited weathering occurring at small land fractions and large CO$_2$ inventories holds for a wide range of parameters. 

\subsubsection{Uncertainties in Model Formulation}

One important caveat to the result that there is only one steady-state solution that planets evolve to from \S \ref{sec:ic} is that a constant albedo was used in this study. Introducing a variable albedo, capable of capturing the ice-albedo feedback, could lead to a situation where a planet oscillates back and forth between ice covered and ice free states. Such oscillations can occur during transient periods of anomalously high weathering rates \citep{Mills2011}, when degassing rates are low \citep{Kadoya2014}, or when luminosity is low \citep{Menou2015}. However, even in these cases hysteresis caused by different initial atmospheric CO$_2$ concentrations is not expected, because the ice free-ice covered oscillations are caused when the temperature predicted by the steady-state carbon cycle is lower than a threshold value, where runaway glaciation occurs. Thus, the presence of an oscillatory state does not affect the way a planet evolves into that state from different initial atmospheric CO$_2$ conditions, and the arguments from \S \ref{sec:ic} that hysteresis does not occur still hold. 

Another important source of uncertainty is the climate parameterization, which is a simple approximation of the results of more sophisticated radiative-convective atmosphere models. A more advanced climate model would likely change the details of the results of this study, both in terms of the exact temperatures and atmospheric CO$_2$ concentrations calculated in \S \ref{sec:results}, and the exact position of the weathering regime boundary in $f_{land}-C_{tot}$ space. In particular, the climate parameterization used in this study likely over-predicts surface temperature at very high atmospheric CO$_2$ levels (e.g. above $\sim 10$ bar). Thus the exact temperatures for planets in the supply-limited weathering regime, or the initial temperatures predicted for planets with CO$_2$ rich primordial atmospheres, may be different from what is shown here. However, these differences do not impact the key results in any major way: the ability of the long-term carbon cycle to stabilize climate is assessed using models at much lower temperatures, where equation \eqref{temp} is more accurate, and the transition between supply-limited and kinetically-limited weathering also occurs at relatively low temperatures. Finally, the lack of hysteresis caused by different initial atmospheric CO$_2$ conditions does not depend on the details of the climate parameterization. 

\subsection{Comparison to Previous Studies on Weathering with Small Land Areas} 
\label{sec:comp_previous}

The topic of silicate weathering and climate stabilization on ocean dominated planets is relatively new. \cite{Abbot2012} studied this topic using a similar carbon cycle model, albeit without a treatment for supply-limited weathering, and found that a habitable, stable climate can still be maintained on an Earth-sized planet with an Earth-like CO$_2$ budget at land fractions as low as 10$^{-2}$. My results show that this main conclusion of \cite{Abbot2012} holds even when supply-limited weathering is included, as long erosion rates are higher than $\sim 0.1$ mm yr$^{-1}$ (see Figure \ref{fig:param_test}A). Furthermore, my results extend the study of climate stabilization via the long-term carbon cycle to lower land fractions and varying planetary CO$_2$ inventories, finding that land fractions must be very low, or CO$_2$ budgets much larger than Earth's, for the carbon cycle to be unable to stabilize planetary climate. My results also provides some additional constraints on the possibility of waterworld self-arrest, as proposed by \cite{Abbot2012}, by showing how the weathering timescale varies with both land fraction and atmospheric CO$_2$ content. The case with $T=400$ K (see Figure \ref{fig:tau_w}) is most comparable to the situation considered by \cite{Abbot2012} as the atmospheric CO$_2$ levels are similar, and shows a consistent weathering timescale of 1-10 Myr for a land fraction greater than 10$^{-2}$.  Thus supply-limited weathering does not significantly impact the weathering timescale, again as long as erosion rates are high. Atmospheric CO$_2$ content, however, is important as the longer weathering timescales that result from higher surface temperatures, and thus higher $P_{CO_2}$, attest to. 

\subsection{Importance of Plate Tectonics for Silicate Weathering and Planetary Habitability}
\label{sec:pt_weathering}

Plate tectonics has long been considered to be an important, and perhaps even necessary, feature of a habitable planet, because plate tectonics facilitates the long-term carbon cycle \citep[e.g.][]{Walker1981,Gonzalez2001,Kasting2003}. Plate tectonics rejuvenates the supply of weatherable rock at the surface through uplift and orogeny, and seafloor production at mid-ocean ridges. Plate tectonics also allows for the recycling of carbon into the mantle, which sustains CO$_2$ degassing to the atmosphere at both arcs and ridges. By including the effects of supply-limited weathering, this study directly illustrates the importance of plate tectonics in planetary habitability, through the influence of plate tectonics on erosion rates. The maximum erosion rate, which controls the supply limit to weathering, is related to tectonics because over long time scales erosion can only proceed as quickly as tectonics brings rock to the surface. Uplift and volcanic resurfacing therefore provide the ultimate, long-term upper bound on erosion rates. As orogeny, uplift, and volcanism are essential aspects of plate tectonics, planets with plate tectonics should have higher maximum erosion rates, and therefore smaller supply-limited weathering regimes. In other words, planets with plate tectonics are more likely to have a carbon cycle that regulates climate, and are therefore more likely to be habitable, because higher erosion rates mean kinetically-limited weathering can be sustained at smaller land fractions and larger planetary CO$_2$ inventories. Furthermore, as plate tectonics is primarily responsible for the formation of continental crust on Earth via arc volcanism, plate tectonic planets may also have higher land fractions compared to stagnant lid planets. 

\section{Conclusions}  
\label{sec:conclusions}

A simple coupled plate tectonic-carbon cycle model shows that the ability of the long-term carbon cycle to buffer climate in response to changes in solar luminosity is preserved when the dependence of plate tectonics on surface temperature, based on recent geodynamical studies, is included.  The climate feedbacks involved in silicate weathering far outweigh the influence climate has on plate tectonics. Moreover, the influence of surface temperature on plate speed actually enhances climate buffering, because warmer temperatures lead to slower plate speeds and lower CO$_2$ degassing rates, in addition to higher weathering rates.  Furthermore, initial conditions are found to not influence the final state reached when the carbon cycle comes to steady-state.  As long as liquid water is present and silicate weathering can occur, a planet with a given exposed land area, total CO$_2$ budget, and receiving a given solar irradiance will evolve to the same surface temperature, regardless of initial atmospheric CO$_2$ concentrations. An initially hot, CO$_2$ rich atmosphere is not an obstacle for developing a temperate climate and active plate tectonics on a planet.

However, the area of exposed land and total CO$_2$ budget of a planet both influence whether the steady-state climate that planet evolves to is clement. At low land areas and high CO$_2$ budgets continental weathering becomes supply-limited, meaning the supply of bedrock to the surface via erosion controls the weathering rate, rather than the kinetics of the reaction between CO$_2$ and silicate minerals. When weathering is supply-limited, it no longer depends on atmospheric CO$_2$ concentration, and the climate feedback is lost. Thus, planets that lie in the supply-limited weathering regime have hot, CO$_2$ rich climates and lack the weathering feedbacks necessary to draw down atmospheric CO$_2$ concentrations in response to increases in solar luminosity. The hot climates that result from supply-limited weathering will cause rapid water loss. Whether the oceans are completely lost, however, likely depends on the amount of CO$_2$ in the atmosphere of a planet in the supply-limited regime. Planets lying near the edge of the supply-limited regime have lower atmospheric CO$_2$ concentrations and can probably reestablish temperate, stable climates as water loss exposes more land and enhances terrestrial weathering.  However, planets lying deep within the supply-limited regime have much larger atmospheric CO$_2$ concentrations and therefore may undergo total water loss before weathering can re-stabilize the climate. Thus ocean dominated planets with CO$_2$ inventories significantly greater than Earth's, such that they would lie well within the supply-limited regime, are less likely to be habitable. High erosion rates are important for preventing supply-limited weathering on a planet because they enhance the supply of bedrock to the surface. The role of erosion rate further highlights the importance of plate tectonics for planetary habitability: plate tectonics drives uplift and orogeny, and therefore plays a direct role in maintaining high erosion rates on a planet.



\acknowledgments

I thank Josh West for discussions on supply-limited weathering, John D. Platt for help with mathematical analysis of the steady-state solutions, and John Chambers and Peter Driscoll for comments on an earlier version of the manuscript. I also thank Dorian Abbot for a thoughtful and constructive review that helped to improve the manuscript. This work was supported by the NASA Astrobiology Institute under cooperative agreement NNA09DA81A. 

\clearpage

\appendix
\section{Appendix}

\setcounter{figure}{0}
\renewcommand{\thefigure}{A\arabic{figure}}

\subsection{Weathering Flux Derivative}
\label{sec:derivative}

The derivative of $F_{weather}/v$ with respect to $R_{oc}$, as introduced in equation \eqref{Rm_Rp}, is derived in this section. Using equations \eqref{my_fw} and \eqref{weather}, 
\begin{equation}
\eqlbl{fw_deriv}
\frac{\partial}{\partial R_{oc}} \left(\frac{F_{weather}}{v} \right) = \frac{\partial v^{-1}}{\partial R_{oc}} F_{weather} - \frac{F_{w_s}}{v} \frac{\partial}{\partial R_{oc}} \left(\exp{\left(-\frac{F_{w_k}}{F_{w_s}} \right)} \right) ,
\end{equation}
where 
\begin{equation} 
\eqlbl{dv_droc}
\frac{\partial v^{-1}}{\partial R_{oc}} = \frac{1.6 v_s k_c M_{H_2O}}{v^2(R_{oc}+M_{H_2O})^2P_{CO_2}^*} \left(\frac{P_{CO_2}}{P_{CO_2}^*} \right)^{-0.654}
\end{equation}
and $v_s = 0.011$ cm yr$^{-1}$ K$^{-1}$ (see equation \eqref{plate_speed}). Evaluating the second term on the right hand side of equation \eqref{fw_deriv}, 
\begin{equation} 
\begin{split}
\eqlbl{dfw_droc}
\frac{\partial}{\partial R_{oc}} \left(\exp{\left(-\frac{F_{w_k} }{F_{w_s} } \right)} \right) & = -\exp{\left(-\frac{F_{w_k} }{F_{w_s}} \right)} \frac{F_{w_k}  M_{H_2O} k_c}{F_{w_s} P_{CO_2}^* (R_{oc}+M_{H_2O})^2} \times \\ & \left[\beta \left(\frac{P_{CO_2}^*}{P_{CO_2}}\right) + 1.6 \left(\frac{P_{CO_2}}{P_{CO_2}^*}\right)^{-0.654} \left(\frac{E_a + \bar{m}_w L_w a}{RT^2}\right) \right] .
\end{split}
\end{equation} 
Combining equations \eqref{fw_deriv}, \eqref{dv_droc}, and \eqref{dfw_droc} shows that the derivative of $F_{weather}/v$ with respect to $R_{oc}$ is positive, as described in the main text (\S \ref{sec:ic}). 

\subsection{Linear Stability Analysis}
\label{sec:stability_analysis}

The stability of the steady-state solution for a given planetary CO$_2$ budget, land fraction, and solar luminosity can be assessed using a linear stability analysis. Although the global carbon cycle model involves three differential equations, there are only two independent equations because one reservoir can always be related to the other reservoirs via the planetary CO$_2$ budget: $C_{tot} = R_{atm} + R_{oc} + R_p + R_{man}$. I chose to treat $R_p$ and $R_{man}$ as the independent dimensions, and calculate the atmosphere and ocean reservoirs from $R_{atm} + R_{oc} = C_{tot} - R_p -R_{man}$. The stability of the resulting two dimensional system of ordinary differential equations is tested using the Jacobian matrix, $J$,  
\begin{equation}
\eqlbl{rman_eq_stab}
\frac{\mathrm{d} R_{man}}{\mathrm{d}t} = (1-f)F_{sub} - F_{degas} = f(R_p,R_{man})
\end{equation}
\begin{equation}
\eqlbl{rp_eq_stab}
\frac{\mathrm{d} R_p}{\mathrm{d}t} = \frac{F_{weather}}{2} + F_{sfw} - F_{sub} = g(R_p,R_{man})
\end{equation}
\begin{equation}
J = \begin{bmatrix} \frac{\partial f}{\partial R_{man}}&\frac{\partial f}{\partial R_{p}} \\ \frac{\partial g}{\partial R_{man}}&\frac{\partial g}{\partial R_{p}} \end{bmatrix}_{(\hat{R}_p, \hat{R}_{man})} 
\end{equation}
where the derivatives of $f$ and $g$ are evaluated at $\hat{R}_p$ and $\hat{R}_{man}$, the steady-state values of the plate and mantle reservoirs, respectively (hats denote a quantity at steady-state throughout this section). If the trace of $J$ is negative and the determinant is positive, then the steady-state solution is stable \citep{Strogatz1994}. All terms in $f$ and $g$ that are functions of $P_{CO_2}$ or $T$ are effectively functions of $R_p$ and $R_{man}$ through the planetary CO$_2$ budget; i.e. perturbations in $R_p$ and $R_{man}$ also give rise to perturbations in $R_{atm}$ and $R_{oc}$ in order to keep the planet's CO$_2$ budget balanced. The atmospheric carbon reservoir is written in terms of $R_p$ and $R_{man}$ as $R_{atm} + R_{oc} = C_{tot} - R_p -R_{man}$, where 
\begin{equation} 
\frac{R_{atm} \bar{m}_{CO_2}g}{A_{earth}} = \frac{k_c R_{oc}}{R_{oc} + M_{H_2O}} 
\end{equation}   
from Henry's law. The amount of CO$_2$ dissolved in the ocean is typically small compared to the number of moles of H$_2$O, meaning $R_{oc} << M_{H_2O}$; using this approximation
\begin{equation} 
\eqlbl{ratm}
R_{atm} = \left(1 + \frac{g M_{H_2O} \bar{m}_{CO_2}}{k_c A_{earth}} \right)^{-1} (C_{tot} - R_p - R_{man}) \equiv \frac{C_{tot} - R_p - R_{man}}{B}.
\end{equation}

With these definitions, 
\begin{equation} 
\begin{split}
\frac{\partial f}{\partial R_{man}} \bigg|_{\hat{R}_p, \hat{R}_{man}} = & \left(\frac{\hat{R}_p L (1-f)}{A_p} - \frac{\hat{R}_{man} f_d 2 L d_{melt}}{V_{man}}\right) \frac{\partial v}{\partial R_{man}} \bigg|_{\hat{R}_p, \hat{R}_{man}} - \\ & \frac{f_d 2 L \hat{v} d_{melt}}{V_{man}} = - \frac{f_d 2 L \hat{v} d_{melt}}{V_{man}} 
\end{split}
\end{equation}
because $F_{sub}(1-f) = F_{degas}$ at steady-state, and  
\begin{equation} 
\frac{\partial f}{\partial R_{p}} \bigg|_{\hat{R}_p, \hat{R}_{man}} = \left(\frac{\hat{R}_p L (1-f)}{A_p} - \frac{\hat{R}_{man} f_d 2 L d_{melt}}{V_{man}}\right) \frac{\partial v}{\partial R_{p}} \bigg|_{\hat{R}_p, \hat{R}_{man}} + \frac{\hat{v}L(1-f)}{A_p} = \frac{\hat{v}L(1-f)}{A_p} . 
\end{equation}
Therefore $\partial f /\partial R_{man}$ is guaranteed to be negative, and $\partial f /\partial R_p$ is positive. The derivatives of $g$ are more complicated due to the seafloor and terrestrial weathering fluxes, but can be written as 
\begin{equation} 
\begin{split}
\eqlbl{dg_drp}
\frac{\partial g}{\partial R_{p}} \bigg|_{\hat{R}_p, \hat{R}_{man}} = & - \frac{L}{A_p} \left(\hat{v} + \hat{R}_p \frac{\partial v}{\partial R_p} \bigg|_{\hat{R}_p, \hat{R}_{man}} \right) + \left(\frac{1}{2} \right)\frac{\partial F_{weather}(P_{CO_2},T)}{\partial R_p} \bigg|_{\hat{R}_p, \hat{R}_{man}} + \\ & \frac{\partial F_{sfw}(P_{CO_2},v)}{\partial R_p} \bigg|_{\hat{R}_p, \hat{R}_{man}}
\end{split}
\end{equation}
and 
\begin{equation} 
\begin{split}
\eqlbl{dg_drman}
\frac{\partial g}{\partial R_{man}} \bigg|_{\hat{R}_p, \hat{R}_{man}} = & - \frac{\hat{R}_p L}{A_p} \left(\frac{\partial v}{\partial R_{man}} \right)_{\hat{R}_p, \hat{R}_{man}}+ \left(\frac{1}{2} \right)\frac{\partial F_{weather}(P_{CO_2},T)}{\partial R_{man}} \bigg|_{\hat{R}_p, \hat{R}_{man}} + \\ & \frac{\partial F_{sfw}(P_{CO_2},v)}{\partial R_{man}} \bigg|_{\hat{R}_p, \hat{R}_{man}} ,
\end{split}
\end{equation}
where 
\begin{equation} 
\eqlbl{dc_dr}
\frac{\partial P_{CO_2}}{\partial R_p} = \frac{\partial P_{CO_2}}{\partial R_{man}} = -\frac{g \bar{m}_{CO_2}}{A_{earth}B}
\end{equation}
\begin{equation} 
\eqlbl{dt_dr}
\frac{\partial T}{\partial R_p} = \frac{\partial T}{\partial R_{man}} = \left( \frac{1.6}{P_{CO_2}^*} \right ) \left(\frac{P_{CO_2}}{P_{CO_2}^*} \right)^{-0.654} \frac{\partial P_{CO_2}}{\partial R_p}
\end{equation}
\begin{equation} 
\eqlbl{dv_dr}
\frac{\partial v}{\partial R_p} = \frac{\partial v}{\partial R_{man}} = -v_s \frac{\partial T}{\partial R_p} .
\end{equation}
Atmospheric CO$_2$ decreases with increasing $R_p$ or $R_{man}$, because increasing the amount of carbon in the plate or mantle reservoirs leads directly to a decrease in the atmospheric and oceanic carbon reservoirs in order to keep the CO$_2$ budget balanced. As a result, temperature also decreases with increasing $R_p$ or $R_{man}$, and plate speed increases due to the drop in surface temperature. As changing either the plate or mantle carbon reservoirs causes a corresponding change in the atmospheric reservoir, $R_{atm}$ is a linear function of both $R_p$ and $R_{man}$ with the same slope (see equation \eqref{ratm}), and $\partial P_{CO_2}/\partial R_p = \partial P_{CO_2}/\partial R_{man}$, $\partial T/\partial R_p = \partial T/\partial R_{man}$, and $\partial v/\partial R_p = \partial v/\partial R_{man}$. 

Using equations \eqref{dc_dr}-\eqref{dv_dr}, the derivatives of the terrestrial and seafloor weathering fluxes can be determined. As derivatives of $P_{CO_2}$, $T$, and $v$ with respect to $R_p$ are equal to those with respect to $R_{man}$, the derivatives of $F_{weather}$ and $F_{sfw}$ with respect to $R_p$ are also equal to those with respect to $R_{man}$. Evaluating those derivatives gives 
\begin{equation}
\eqlbl{dfsfw_dr}
\frac{\partial F_{sfw}}{\partial R_p} = \frac{\partial F_{sfw}}{\partial R_{man}} = \frac{F_{sfw} g \bar{m}_{CO_2}}{v P_{CO_2}^*A_{earth}B} \left[v_s1.6\left(\frac{P_{CO_2}}{P_{CO_2}^*}\right)^{-0.654} - \alpha v \left(\frac{P_{CO_2}^*}{P_{CO_2}}\right) \right]
\end{equation}
and 
\begin{equation}
\begin{split}
\eqlbl{dfw_dr}
\frac{\partial F_{weather}}{\partial R_p} = & \frac{\partial F_{weather}}{\partial R_{man}} = -\exp{\left(-\frac{F_{w_k} }{F_{w_s}} \right)} \frac{F_{w_k} g\bar{m}_{CO_2}}{ P_{CO_2}^* A_{earth} B} \times \\ & \left[\beta \left(\frac{P_{CO_2}^*}{P_{CO_2}}\right) + 1.6 \left(\frac{P_{CO_2}}{P_{CO_2}^*}\right)^{-0.654} \left(\frac{E_a + \bar{m}_w L_w a}{RT^2}\right) \right].
\end{split}
\end{equation}

The fact that plate speed increases with increasing $R_p$ or $R_{man}$ ensures that the first term on the right hand side of both equations \eqref{dg_drp} and \eqref{dg_drman} is negative; a positive perturbation in the plate reservoir will increase the subduction flux both because the plate reservoir is larger, and the plate speed is higher, thereby acting to deplete the plate reservoir and damp out the perturbation.  Likewise, a positive perturbation in the mantle reservoir also increases the plate speed and hence acts to decrease the plate reservoir. The terrestrial weathering flux decreases with increasing $R_p$ or $R_{man}$ (see equation \eqref{dfw_dr}) because both $P_{CO_2}$ and $T$ are lowered by an increase in either the plate or mantle carbon reservoirs. As a result, the second term on the right hand side of equations \eqref{dg_drp} and \eqref{dg_drman} is also negative. There are two competing effects that determine the sign of $\partial F_{sfw}/\partial R_p$ (see equation \eqref{dfsfw_dr}); increasing $R_p$ drops $P_{CO_2}$, acting to decrease $F_{sfw}$, but also increases $v$, acting to increase $F_{sfw}$. The influence of decreasing atmospheric CO$_2$ tends to dominate, so $\partial F_{sfw}/\partial R_p$ is found to be negative for all parameter ranges explored in this study.  

\begin{figure}
\includegraphics[scale = 0.5]{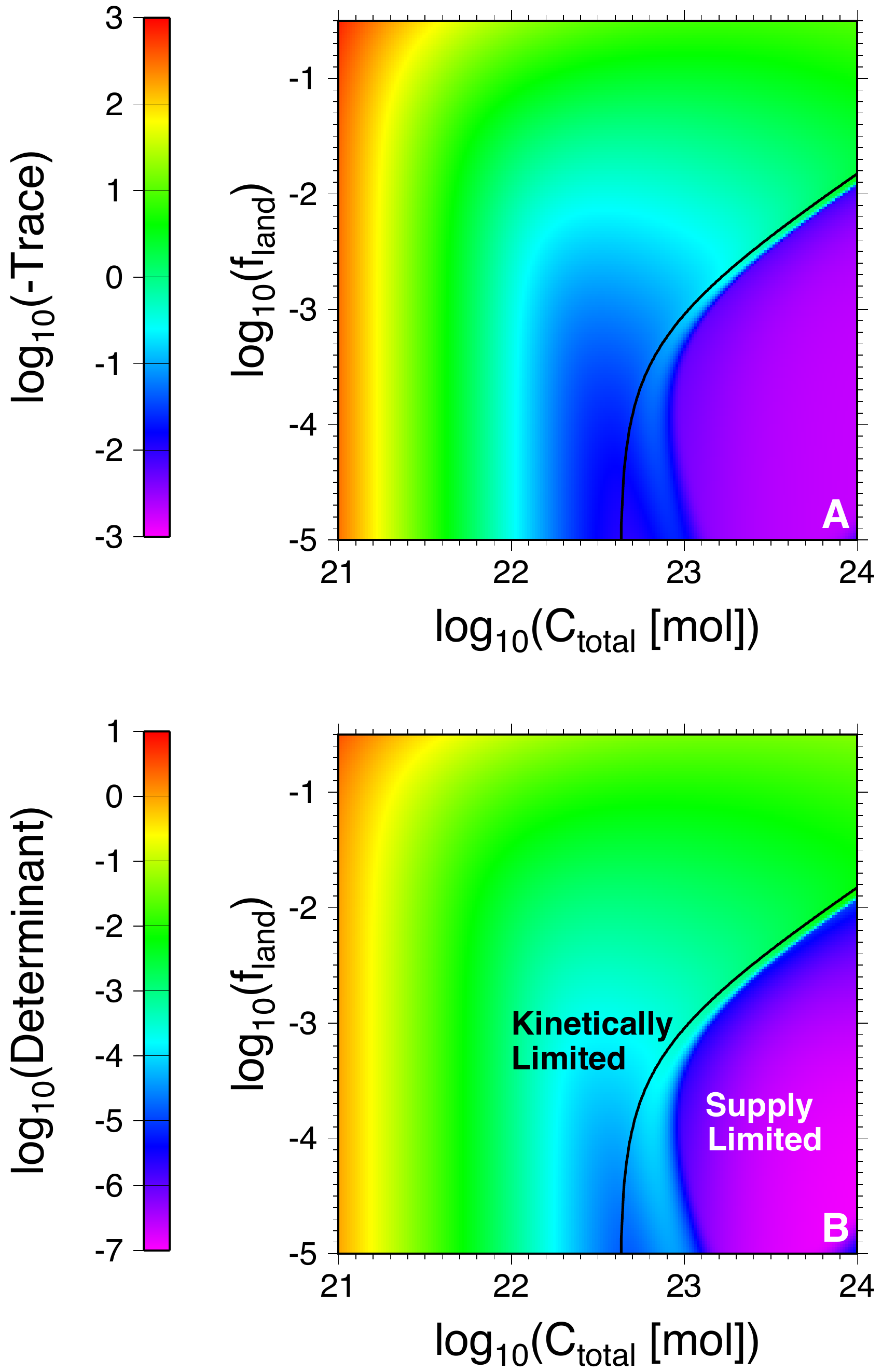}
\caption{\label {fig:stability} Logarithm of the negative of the trace of the Jacobian, $J$, (a) and logarithm of the determinant of $J$ (b) for steady-state solutions to equations \eqref{rp_eq}-\eqref{raoc_eq} as a function of $C_{tot}$ and $f_{land}$. A present day solar luminosity is assumed, and all other parameters are listed in Tables \ref{tab_param} \& \ref{tab_cons}. }  
\end{figure} 

Summing all of the terms, $\partial g/\partial R_p$ and $\partial g/\partial R_{man}$ are negative.  Thus the trace of $J$, $\partial f/\partial R_{man} +\partial g/\partial R_p$, is negative, and the determinant, $(\partial f/\partial R_{man})(\partial g/\partial R_p)-(\partial g/\partial R_{man})(\partial f/\partial R_p)$, is positive, and steady-state solutions are stable. Figure \ref{fig:stability} shows the logarithm of the negative of the trace of $J$ and the logarithm of the determinant of $J$ for steady-state solutions to equations \eqref{rp_eq}-\eqref{raoc_eq} over a wide range of $C_{tot}$ and $f_{land}$. Though the trace and determinant approach 0 at large $C_{tot}$ and small $f_{land}$, particularly in the supply-limited weathering regime (see \S \ref{sec:fland}), the trace remains negative and determinant remains positive, demonstrating that these steady-states are stable.   

\clearpage

\def\els{/Users/bfoley/Documents/tex/elsart}
\def\agu{/Users/bfoley/Documents/tex/agu}
\def\bib{/Users/bfoley/Documents/tex/bib}

\bibliographystyle{model5-names}
\bibliography{GeneralGeophys+Misc,MantleConvection,TreatiseXtra,seismology,TwoPhase-Damage,planetaryscience}

\end{document}